\documentclass[12pt]{iopart}
\usepackage{graphicx}

\expandafter\let\csname equation*\endcsname\relax
\expandafter\let\csname endequation*\endcsname\relax
\usepackage{amsmath,amssymb}  
\usepackage{bm}  
\usepackage{cite} 
\usepackage[titletoc,toc,title]{appendix}
\usepackage{graphicx}
\usepackage{subfig}
\usepackage{caption}
\usepackage{sidecap}

\begin{document}

\title{A statistical physics approach to learning curves for the Inverse Ising problem}
\author{Ludovica Bachschmid-Romano
and Manfred Opper}
\address{Department of Artificial Intelligence, Technische Universit\"{a}t Berlin,
Marchstra{\ss}e 23, Berlin 10587, Germany}
\ead{ludovica.bachschmidromano@tu-berlin.de and manfred.opper@tu-berlin.de}

\begin{abstract}
Using methods of statistical physics, we analyse the error of learning couplings in large Ising models from independent data (the inverse Ising problem). We concentrate on
learning based on local cost functions, such as the pseudo--likelihood method for which the couplings  are inferred independently for each spin. Assuming that the data are generated
from a true Ising model, we compute the reconstruction error of the couplings using
a combination of the replica method with the cavity approach for densely connected
systems. We show that an explicit estimator based on a quadratic cost function
achieves minimal reconstruction error, but requires the length of the true coupling vector as prior knowledge.
A simple mean field estimator of the couplings which does not need such knowledge is asymptotically optimal, i.e.
when the number of observations is much large than the number of spins. 
Comparison of the theory with numerical simulations shows excellent agreement for data generated from two models with random couplings in the high temperature 
region: a model with independent couplings (Sherrington--Kirkpatrick model), and a model where the matrix of couplings has a Wishart distribution.
\end{abstract}

\section{Introduction}
In recent years, there has been an increasing interest in applying classical Ising models
to data modelling. Applications range from modelling the dependencies of
spikes recorded from ensembles of neurons \cite{Bialek,roudi2009ising} to protein structure determination \cite{Weigt} or gene expression analysis \cite{Lezon}. An important issue for such applications is the so--called inverse Ising problem, i.e. the statistical problem of
fitting model parameters, external fields and couplings, to a set of data. Unfortunately, the 
exact computation of statistically efficient estimators 
such as the {\em maximum likelihood} estimator is computationally intractable for large systems.
Hence, to overcome this problem researchers have suggested two possible solutions: the first one
tries to approximate maximum likelihood estimators by computationally efficient procedures such as Monte Carlo
sampling  \cite{MC} or mean field types of analytical computations, see e.g. \cite{Kappen,tanaka1998mean,Roudi2009,sessak}.
 A second line of research abandons the
idea of maximising the likelihood function and replaces it by other cost functions which are easier to optimise. The most 
prominent example is the so--called pseudo--likelihood method \cite{Besag,Aurell,Decelle_Ricci,Mozeika,tyagi}.  In general  it is not clear which of the two methods leads to better reconstruction of an Ising model.
The quality of such estimators, e.g. measured by the mean squared reconstruction error of network parameters, will
depend on the problem at hand.

As an alternative to analysing specific instances of problems, one may study the typical prediction performance of algorithms assuming that the {\em true} Ising parameters are drawn at random from a given ensemble distribution. For such random problem cases, one can apply powerful methods of statistical physics to compute (scaled) reconstruction errors
{\em exactly} in the limit where the number of spins grows to infinity and the number of data 
is increased proportionally to the number of spins.
Such an approach has been applied extensively to statistical learning in large neural networks in the past
\cite{Opper_Kinzel,Engel,nishimori}.
In a previous paper \cite{LBR_MO_15} we have applied this method to the learning from dynamical data 
which are modelled by a kinetic Ising model with random independent couplings. This problem is theoretically simpler compared to the static, 'equilibrium' Ising case 
discussed in the present paper. This is because the spin statistics of the dynamical model is fairly simple
in the 'thermodynamic' limit of a large network and gives rise to Gaussian distributed fields.

We will show in the following that a related approach is possible to data drawn independently from an 
equilibrium Ising model when we assume that couplings are learnt independently for each spin 
using local cost functions. Although the spin statistics is more complicated, computations are possible, when 
the so--called  'cavity'  method \cite{mezard-parisi-virasoro} is applicable to the true teacher Ising model.

The paper is organised as follows: Chapter 2 explains the inverse Ising problem and 
maximum likelihood estimation. Chapter 3  introduces simpler estimators which are derived from local cost functions.
In Chapter 4, we review the statistical physics approach for analysing learning performances within the 
so--called teacher
student scenario. In Chapter 5 we explain the cavity method for performing quenched averages over spin
configurations. Chapter 6 presents explicit results of our method 
applied to the inverse Ising model with independent Gaussian couplings
(SK--model). 
In Chapter 7 we study the learning performance of algorithms based on local quadratic cost functions and we compute the optimal local quadratic cost function. In Chapter 8 we show that an optimal quadratic function provides the best local estimator for the couplings.
Chapter 9 introduces further applications of the cavity method which allow us to 
simplify order parameters  corresponding to the true teacher couplings. As an example, we compute 
 the reconstruction error for an Ising model with Wishart distributed, i.e. weakly dependent couplings.
The method is also applied to re-derive a simple mean field approximation to the 
maximum likelihood estimator.  Chapter 10 explains how the mean field estimator can be obtained from a 
local cost function and presents results for the reconstruction errors. Chapter 11 discusses the asymptotics of
the reconstruction errors for large number of data and relates these results to expressions known from classical statistics. Chapter 12 contains comparisons of our results with those of simulations of the estimators and 
Chapter 13 presents a summary and an outlook.

\section{Estimators for the Inverse Ising model}
Let us consider a system of $N$ binary spin variables $\bm{ \sigma}=(\sigma_0, \dots, \sigma_{N-1})$ connected by pairwise interactions  $J_{ij}$ and subject to external local fields $H_i$.
The  probability distribution of the spin set is given by the Boltzmann equilibrium distribution
\begin{equation}
P( \bm{ \sigma} \vert \bm{J},\bm{H}) = Z_{\mbox{\tiny{Ising}}}^{-1} \exp \left[ \beta \sum_{i<j}J_{ij} \sigma_i \sigma_j+ \beta \sum_i H_i \sigma_i \right],
\label{eq:Ising}
\end{equation}
where $Z_{\mbox{\tiny{Ising}}}$ is the partition function and $\beta$ is the inverse temperature.
Given a set of  $M$ independent observations $\{ \bm{ \sigma}^k \} _{k=1} ^M$ drawn independently 
from  (\ref{eq:Ising}), the inverse Ising problem consists of estimating the  model parameters $\bm{H}$ and  $\bm{J}$
from the data.
A standard approach for parameter estimation is the maximum likelihood (ML) method, 
which has the properties of  consistency  and asymptotic efficiency \cite{schervish2012theory}. 
Maximum likelihood can be formulated as the  minimisation of the following cost function (negative log--likelihood)
\begin{equation}
E_{ML}(\bm{J},\bm{H}) = - \sum_{k=1}^M \ln P( \bm{ \sigma}^k \vert \bm{J},\bm{H}) 
\label{eq:ln-lik}
\end{equation}
with respect to the matrix of couplings $\bm{J}$ and the field vector $\bm{H}$. As is well known, the minimisation
of  (\ref{eq:ln-lik}) is equivalent to a simple set of conditions for the first and second
moments of the ensemble (\ref{eq:Ising}) of spins: the parameters estimated by ML
lead to the matching of the empirical (data averaged) magnetisations 
to the magnetisation given by the model (\ref{eq:Ising}). Likewise
we have the matching of all empirical pair correlations of spins 
with their model counterparts.
Despite the simplicity of this rule, the practical minimisation of (\ref{eq:ln-lik}) requires
the computation of these spin moments for a given set of couplings and fields which 
is equivalent to averaging over $2^N$ spin configurations, which is intractable for larger $N$.
An approximation of such averages by  Monte Carlo sampling is possible but requires sufficient 
time for equilibration. Alternatively, different approximation techniques have been developed to provide a good estimate of the parameters at a smaller computational cost, see e.g. \cite{Roudi2009,sessak,Aurell,Decelle_Ricci,cocco2011,Minimum_prob_flow,ricci2012,nguyen2012bethe}.

\section{Local Learning}
If we neglect the symmetry of coupling matrix, i.e. the equality $J_{ij} = J_{ji}$, we can develop estimators 
which learn the 'ingoing' coupling vectors $J_{ij}$ for $j=0, \ldots, i-1,i+1, \ldots, N-1$ for each spin
$\sigma_i$  independently. It turns out that the corresponding (local) algorithms 
can often be performed in a much more efficient way compared to the ML method.

In the following we will concentrate on the estimation of the couplings only and set the external fields $H_i$ to zero. 
We will specialise on the couplings for spin $\sigma_0$ and 
assuming that the typical couplings ${J}_{ij}$ are variables with magnitude scaling like $1/\sqrt{N}$ 
for large $N$. We define a vector of rescaled couplings (weights) as
\begin{equation}
\bm{W} = (W_1,\ldots,W_{N-1}) \doteq {\sqrt{N}} (J_{01}, \ldots, J_{0 N-1}).
\label{eq:en_fz_0}
\end{equation}
We will assume that an estimator for $\bm{W}$ is defined by the minimisation of a cost function,
\begin{equation}
E(\bm{W})=\sum_{k=1}^M \mathcal{E}(\bm{W}; \bm{\sigma}^k),
\label{eq:en_fz}
\end{equation}
which is additive in the observed data. An important and widely used case is the pseudo--likelihood approach, where 
the cost function 
\begin{equation}
\begin{split}
\mathcal{E}(\bm{W}; \bm{\sigma}) &= - \ln P (\sigma_0 \vert  \bm{ \sigma}_{ \setminus 0}, \bm{W})  \\
& = - \beta \sigma_0 \sum_{j \neq 0} \frac{W_j \sigma_j}{\sqrt{N}}
+ \ln \left(2 \cosh     \beta  \sum_{j \neq 0}  \frac{W_j \sigma_j}{\sqrt{N}} \right)
\label{pseudo-_cost}
\end{split}
\end{equation}
is given by the negative log--probability of spin $\sigma_0$ conditioned
on all other spins $\bm{ \sigma}_{ \setminus 0}$. In contrast 
to the ML approach, the gradient of this function can be computed 
in an efficient way.


\section{Teacher--student scenario and statistical physics analysis }
We assume in the following that data are generated independently at random 
from a 'teacher' network with 
coupling matrix $J_{ij}^*$. A local learning algorithm based on the minimisation of
(\ref{eq:en_fz}) produces
'student' network couplings $\bm{W}$ 
as estimators for the teacher network couplings $\bm{W}^ *=  {\sqrt{N}} (J^*_{01}, \ldots, J^*_{0 N-1})$.
To measure the quality of a given local learning algorithm, we  will compute the 
average square reconstruction error given by
\begin{equation}
\varepsilon= N^{-1} \overline{(\bm{W}^ * - \bm{W})^2} = Y - 2 \rho + Q,
\label{eq:error}
\end{equation}
where we define order parameters
\begin{equation}
Y= N^{-1} \overline{(\bm{W^*})^2},   \qquad
Q= N^{-1} \overline{(\bm{W})^2}, \qquad
\rho= N^{-1} \overline{\bm{W^*} \cdot \bm{W}},\label{eq:def_rho}
\end{equation}
representing, respectively, the squared lengths of the teacher and student coupling vectors and the overlap between  teacher and a student coupling vectors. Here the overline defines an expectation over the ensemble of $M=\alpha N$
training data drawn at random from an Ising model with teacher couplings $\bm{J}^*$, i.e.
\begin{equation}
\overline{(\dots)}= 
\sum_{\bm{\sigma^1,\ldots, \sigma^{M}} }  \prod_{k=1}^M P(\bm{\sigma^k} \vert \bm{J}^ *)  (\dots).
\label{eq:quenched}
\end{equation}
Since there is often no explicit analytical solution to 
the minimisers  $\bm{W}$ of (\ref{eq:en_fz}), we will resort to a statistical physics approach
which has been successfully applied to the analysis of a great variety of
problems related to learning in neural networks \cite{Opper_Kinzel,Engel,nishimori}. In this approach \cite{Seung-Sompolinsky-Tishby}, one defines
 a statistical ensemble of student weights by 
a Gibbs distribution 
 \begin{equation}
p(\bm{W}) = \frac{1}{Z} \exp[- \nu E(\bm{W})],
\label{eq:Gibbs}
\end{equation}
with the partition function 
\begin{equation}
Z= \int \; d \bm{W}  \exp[- \nu E(\bm{W})],
\end{equation}
where $1/\nu$ represents an effective temperature which controls the fluctuations of the 
'training energy' $E(\bm{W})$. Using techniques from statistical physics of
disordered systems one computes order parameters at nonzero temperature 
and performs the limit $\nu\to\infty$ at the end of the calculation. The 'thermal average' 
$\langle \bm{W} \rangle$ with respect to the 
distribution (\ref{eq:Gibbs}) converges to the minimiser of the cost function $E(\bm{W})$. Order parameters
can be extracted from the quenched average of the free energy $F$ corresponding to (\ref{eq:quenched})
using the replica method:
\begin{equation}
F=- N^{-1}\nu^{-1} \overline{ \ln Z } = - \lim_{n \to 0}  N^{-1}\nu^{-1} \frac{\partial}{\partial n}  \ln \overline{ Z^n },
\label{eq:replica_trick}
\end{equation}
where the average replicated partition function for integer $n$ is given by
\begin{equation}
 \overline{ Z^n } = \int   \prod_{a=1}^n  \; d \bm{W}^a     \left\{
 \sum_{\bm{\sigma}}   P(\bm{\sigma}  \vert \bm{J}^ *)    \exp[- \nu  \sum_{a=1}^n \mathcal{E}(\bm{W^a}; \bm{\sigma} )]
 \right\}^{\alpha N}.
\label{eq:Z_n}
\end{equation}
To allow for an analytical treatment, 
we assume that the local cost
 function $\mathcal{E}(\bm{W^a}; \bm{\sigma} )$ depends on the spins and couplings only via 
 $\sigma_0$ and the local field
 $
 h \doteq \frac{1}{\sqrt{N}} \sum_{j\neq 0} W_j \sigma_j
 $
in the following way:
 \begin{equation}
 \mathcal{E}(\bm{W}; \bm{\sigma} ) = \Phi(\sigma_0 h).
 \label{eq:def_Phi}
 \end{equation}
 Obviously, the pseudo--likelihood cost function (\ref{pseudo-_cost}) belongs to this class of functions.
The goal of the following Section is to perform the expectation (\ref{eq:Z_n}). 
The resulting expression depends on a set of order parameters and can for integer $n$ be evaluated by standard saddle--point methods in the limit $N\to\infty$. Performing an analytical continuation for $n\to 0$ yields both the free energy and
the self--averaging values of these order parameters. 
While in most previous applications \cite{Opper_Kinzel,Engel,nishimori} of this programme to learning in neural networks, the 
quenched average over data in (\ref{eq:Z_n}) is straightforward, the required average over 
Ising spin configurations drawn from the distribution (\ref{eq:Ising}) cannot be performed (for 
arbitrary $N$) in closed form. One might attempt a solution to this problem by introducing 
a second set of replicas which would deal with the partition function $Z_{\mbox{\tiny{Ising}}}^{-1}$ in the 
denominator of (\ref{eq:Ising}). We expect that such an approach can be carried on for random teacher 
couplings but 
may lead to complicated expressions which have to be carefully evaluated for $N\to\infty$. 
In the next Section we will use a simpler approach using  ideas of the cavity method \cite{mezard-parisi-virasoro}
which allows, under certain assumptions on the teacher coupling matrix $\bm{J}^*$,  the explicit 
computation of the quenched average for $N\to\infty$.

 \section{Cavity approach I: quenched averages}
In order to perform  the quenched average  in (\ref{eq:Z_n}), we will 
combine the replica approach with ideas of the so--called cavity method. In doing so we write 
the Gibbs distribution (\ref{eq:Ising}) corresponding to the teacher couplings 
in the form 
\begin{equation}
P( \bm{ \sigma} \vert \bm{J}^*) \propto \exp \left[ \beta \sigma_0 \sum_ {j\neq 0} J_{0j}^* \sigma_j  \right] P_{\mbox{\tiny{cav}}} ( \bm{ \sigma} \backslash \sigma_0),
\end{equation}
where $P_{\mbox{\tiny{cav}}}$ denotes the distribution of the remaining spins in a system where the spin $\sigma_0$ 
was {\em removed}, creating a cavity at this site, which gives the method its name. The replicated partition function depends only on the fields
$h_a \doteq \frac{1}{\sqrt{N}} \sum_{j\neq 0} W_j^a \sigma_j$ where $a \in \{*,1,\ldots,n\}$.
The cavity assumption for the statistics of such fields in densely connected systems can be summarised as follows:
in performing expectations over $P_{\mbox{\tiny{cav}}}$, we can assume that dependencies between 
spins are so weak that random variables $h_a$  become jointly Gaussian distributed
in the limit $N\to\infty$.  Hence, 
the joint distribution of spin $\sigma_0$ and the fields can be expressed as
\begin{equation}
P(\sigma_0, h_*, h_1,\ldots h_n) = \frac{1}{Z_0}
e^{\beta \sigma_0 h_*} p_{\mbox{\tiny{cav}}}(h_*,h_1,\ldots h_n)
\end{equation}
with the normalisation
\begin{equation}
Z_0 = 2\int \cosh(\beta h_*) p_{\mbox{\tiny{cav}}}(h_*) dh_* \ .
\end{equation}
Assuming that in absence of external fields we have vanishing magnetisations (paramagnetic phase), 
the distribution $p_{\mbox{\tiny{cav}}}(h_*,h_1,\ldots h_n)$ is a multivariate Gaussian density with zero mean
and covariance
\begin{equation}
\left\langle h_a h_b  \right\rangle = \frac{1}{N}\sum_{i,j\neq 0} W_i^a C^{\backslash 0}_{ij} W_j^b .
\label{eq:cav_h}
\end{equation}
The matrix $C^{\backslash 0}$ is the correlation matrix of the reduced spin system (without $\sigma_0$), which 
does not depend on the couplings $\bm{W}^*$. We have $C^{\backslash 0}_{ii} = 1$ and assume 
that typically $C^{\backslash 0}_{ij} = O(\frac{1}{\sqrt{N}})$ for $i\neq j$ and large $N$.
However, this scaling does  not mean that we can neglect the non--diagonal matrix elements. We will later see that 
they give {\em nontrivial contributions} to the final reconstruction error.  
Within this framework, the quenched average in 
(\ref{eq:Z_n}) is rewritten in terms of integrals over the random variables $h_a$ as follows:
\begin{equation}
\begin{split}
&\sum_{\bm{\sigma}}   P(\bm{\sigma}  \vert \bm{J}^*)    \exp[- \nu  \sum_{a=1}^n \mathcal{E}(\bm{W^a}; \bm{\sigma} )]
 = \\
 & \sum_{\sigma_0} \int  \,  dh_* \prod_{a=1}^n dh_a \,      
\frac {1}{Z_0}  \exp \left[ \beta \sigma_0 h_* \right] 
 \exp \left[- \nu  \sum_{a=1}^n \Phi( \sigma_0 h_a) \right]  \;  
p_{\mbox{\tiny{cav}}}(h_*,h_1,\ldots h_n).
\end{split}
\label{eq:F_inner_average}
\end{equation}
This result can be expressed by the covariances
 (\ref{eq:cav_h}) which in the limit $N\to\infty$ will
become {\em self averaging order parameters} which will be computed by the replica method (Appendix \ref{sec:AppA}).
Under the  assumption of replica symmetry (which is expected to be correct for convex 
cost functions, which holds e.g. in the case of pseudo--likelihood),
these new order parameters and their physical meaning are denoted as:
\begin{equation}
\begin{split}
V   & \doteq \frac{1}{N}\sum_{i,j\neq 0} W_i^* C^{\backslash 0}_{ij} W_j^* \\
R & \doteq \frac{1}{N}\sum_{i,j\neq 0} W_i^* C^{\backslash 0}_{ij} \left\langle W_j\right\rangle_w  = \frac{1}{N}\sum_{i,j\neq 0} W_i^* C^{\backslash 0}_{ij} W_j^a \qquad a  \neq *,\\
q_0  & \doteq \frac{1}{N}\sum_{i,j\neq 0}  \left\langle W_i  W_j \right\rangle_w C^{\backslash 0}_{ij}  =
\frac{1}{N}\sum_{i,j\neq 0} W_i^a C^{\backslash 0}_{ij} W_j^a      \qquad a  \neq *, \\
q & \doteq \frac{1}{N}\sum_{i,j\neq 0}  \left\langle W_i \right\rangle_w C^{\backslash 0}_{ij}  \left\langle W_j \right\rangle_w 
 =   \frac{1}{N}\sum_{i,j\neq 0} W_i^a C^{\backslash 0}_{ij} W_j^b   \qquad a\neq b \neq *,
\label{eq:def_parameters}
\end{split}
\end{equation}
where the brackets $\left\langle\ldots \right\rangle_w$ denote averages with respect to the distribution of couplings (\ref{eq:Gibbs}). 
 

\section{Replica result}
{\label{sec:replica}
Using a replica symmetric ansatz, the computations follow the approach 
summarised in Appendix \ref{sec:AppA}. In the  zero temperature limit $\nu \to \infty$,
the fluctuations of student couplings vanish and we obtain the convergence of the order parameters $q_0\to q$  with the limiting 'susceptibility' 
 $$
 x \doteq \lim_{\nu\to\infty} (q_0 - q) \nu =  \lim_{\nu\to\infty}  \frac{\nu}{N}\sum_{i,j\neq 0}
 \left(\left\langle W_i W_j \right\rangle_w - \left\langle W_i \right\rangle_w \left\langle W_j \right\rangle_w 
 \right) C^{\backslash 0}_{ij} 
 $$ 
 remaining finite and nonzero. As a  main result, 
 we find that the auxiliary order parameters (\ref{eq:def_parameters}) are obtained by extremizing the limiting free energy 
function
\begin{equation}
F= -  \underset{q,R,x}{\mbox{extr}} \,  \left\{ \frac{1}{2}  \frac{q - R^2/V}{x}  
 +   \alpha  \int d v  \; G_{\beta R,q} (v) \;\max_y \left[ -\frac{(y-v)^2}{2 x} - \Phi(y) \right] \right\},
  \label{eq:free_energy}
\end{equation}
where $G_{\mu, \omega} (v)$ denotes a
 scalar Gaussian density
 with mean $\mu$ and standard deviation $\omega$.
Remarkably, this free energy does (for any fixed cost function $\Phi$) only depend on the 
teacher couplings $\bm{J}^*$ via the order parameter $V$, defined in equation (\ref{eq:def_parameters}).
To compute the prediction error, however, we need the 'original' order parameters (\ref{eq:def_rho}).
These can be expressed by the auxiliary ones
$q$, $R$ and $x$. This relation can be derived from the free energy (Appendix \ref{sec:AppB}) in a standard way
by adding corresponding external fields to the 'Hamiltonian' in the Gibbs free energy (\ref{eq:Gibbs}).
This relation brings back further statistics related to the teacher couplings $\bm{J}^*$ via
\begin{eqnarray}
&\rho  &=  \frac{R  Y}{V}, \nonumber \\
&Q&=  (q-\frac{R^2 }{V}) \frac{1}{N} \mbox{Tr} \overline{\bm{C}^{-1}} + \frac{R^2  Y}{V^2},
\label{eq:def_rho_Q}
\end{eqnarray}
with the corresponding reconstruction error 
\begin{equation}
\varepsilon =(q-\frac{R^2 }{V}) \frac{1}{N} \mbox{Tr} \overline{\bm{C}^{-1}} + Y(1- \frac{R }{V})^2.
\label{eq:error_def_formula}
\end{equation}
In deriving these results, we have also assumed that for $N\to\infty$, 
$\frac{1}{N} \mbox{Tr} \overline{(\bm{C}^{\backslash 0})^{-1}} \to\frac{1}{N} \mbox{Tr} \overline{(\bm{C})^{-1}}$.
Note that the prediction error is larger than the one we would get if we had neglected the off--diagonal elements of the correlation matrix $\bm{C}^{\backslash 0}$.
The error (\ref{eq:error_def_formula}) depends on the teacher couplings $\bm{J}^*$ through the parameter $Y$ and the parameter $V$ (the
cavity variance of the teacher field)  and through the trace of the inverse correlation matrix $\bm{C}$ corresponding to the
teacher's spin distribution. We will show later that the latter quantity can be expressed by the former using 
a second application of the cavity method. In the next section, we will see that the parameter $V$ can be estimated
from the data.

We will illustrate the result (\ref{eq:error_def_formula}) for the case of random teacher couplings $J_{ij}^*$ drawn independently for $i<j$ from a Gaussian density of variance $1$. This corresponds to the celebrated
{\em Sherrington--Kirkpatrick} (SK) model \cite{SK}. For $\beta<1$, i.e. outside of the spin--glass phase,
our simple form of the cavity arguments are known to be correct \cite{mezard-parisi-virasoro}
and one finds the values 
\begin{equation}
\begin{split}
 &V = Y = 1, \\
 &\lim_{N\to\infty} \frac{1}{N} \mbox{Tr} \overline{(\bm{C})^{-1}} = 1 + \beta^2,
 \label{SK_inverse}
 \end{split}
 \end{equation}
 for zero magnetisations $m_i =0$ in the literature \cite{Bray}.  A comparison of the theory (\ref{eq:error_def_formula}) with  numerical simulations is shown in Section \ref{sec:results}.

 \section{Quadratic cost functions}
\label{sec:linear}
Among the simplest functions satisfying the property (\ref{eq:def_Phi}), we
consider quadratic cost functions of the form
\begin{equation}
E_{\eta}(\bm{W})= \frac{1}{2} \sum_{i\neq 0,j \neq 0}W_{i} \hat{C}_{ij} W_j -\eta \sqrt{N} \sum_{j\neq 0} \hat{C}_{0j} W_j,
\label{eq:cost_fz_quadratic}
\end{equation}
where the empirical correlation matrix is defined as
 \begin{equation}
 \hat{C}_{ij} \doteq \frac{1}{M} \sum_{k=1}^{M} \sigma_i^k \sigma_j^k.
\end{equation} 
These allow for an explicit computation of the estimator in terms of a matrix inversion.
The estimator minimizing  (\ref{eq:cost_fz_quadratic}) is given by 
\begin{equation}
W^\eta_{i} =  
\eta \sqrt{N} \sum_{j\neq 0} (\hat{C}_{-0}^{-1})_{ji} \hat{C}_{0j} \qquad i\neq 0,
\label{eq:lin_Est}
 \end{equation}
where the matrix $\hat{C}_{-0}$ is the submatrix of $\hat{C}$ where the $0$-th column and $0$-th row are deleted (not to be confused with the cavity matrix $C^{\backslash 0}$) and $\eta$ is a free parameter.
The estimation error  can be computed from the free energy (\ref{eq:free_energy}) by setting 
\begin{equation}
\Phi(h) =  \frac{h^2 }{2}  -  \eta h,
\label{eq:quad_cost}
\end{equation}
 and gives (see Appendix \ref{sec:AppC})
\begin{equation}
\varepsilon= \left( \frac{\beta \eta}{1+\beta^2 V}-1 \right)^2 Y   +   \frac{\eta^2 }{(\alpha-1)(1+\beta^2 V)}  \frac{1}{N} Tr 
\overline{\bm{C}^{-1}}.
\label{eq:error_quad}
\end{equation}
The optimal choice for the quadratic cost function (\ref{eq:cost_fz_quadratic}) is found by fixing the parameter 
 $\eta$ to the value that minimizes the error (\ref{eq:error_quad}), namely
\begin{equation}
\eta_{\mbox{opt}}= \frac{(\alpha-1)(1+\beta^2 V) \beta Y}{(\alpha-1) \beta^2 Y +(1+\beta^2 V) \frac{1}{N} Tr 
\overline{\bm{C}^{-1}}  },
\label{eq:eta_opt}
\end{equation}
with the corresponding minimal error
\begin{equation}
\varepsilon_{\mbox{opt}}=  \frac{ (1+\beta^2 V) Y \frac{1}{N} Tr \overline{\bm{C}^{-1}} }{(\alpha-1) \beta^2 Y +(1+\beta^2 V) \frac{1}{N} Tr 
\overline{\bm{C}^{-1}}}  .
\label{eq:err_opt}
\end{equation}
In general, the computation of the optimal parameter $\eta_{\mbox{opt}}$ requires the knowledge
of the three parameters $Y$, $V$ and  $\frac{1}{N} Tr \overline{\bm{C}^{-1}}$ which characterise 
the statistical ensemble to which he unknown teacher matrix $\bm{J}^*$ belongs. However, (\ref{eq:eta_opt}) simplifies as $\alpha\to\infty$ and we get
\begin{equation}
\lim_{\alpha\to\infty} \eta_{\mbox{opt}}  = \frac{1+\beta^2 V}{\beta}.
\label{eq:eta_opt_asym}
\end{equation}
We will now show that the remaining parameter $V$ can be estimated from the observed data.
We use the fact that at its minimum, the cost function (\ref{eq:cost_fz_quadratic}) equals
\begin{equation}
\begin{split}
E_{\eta}(\bm{W}^{\eta})= - \frac{ N}{2} \eta^2 \Delta  ,
 \end{split}
 \label{E_phi}
 \end{equation}
where we have used (\ref{eq:lin_Est}) and defined 
\begin{equation}
\Delta=\sum_{i\neq 0,j\neq 0} \hat{C}_{0i} (\hat{C}_{-0}^{-1})_{ij} \hat{C}_{0j},
\label{eq:Delta}
 \end{equation}
which only depends on the spin data. On the other hand  in the situations where our statistical physics formalism applies, the minimal training energy (\ref{E_phi}) will be self--averaging in the thermodynamic limit $N\to\infty$
and can be computed as the {\em zero temperature limit of the free energy}, i.e. the free energy function 
(\ref {eq:free_energy}) evaluated at the stationary values of the order parameters. The calculation in 
(Appendix \ref{sec:AppC}) yields
\begin{equation}
\Delta= \frac{1+\alpha \beta^2 V}{\alpha (1+\beta^2 V)}.
\label{delta_explicit}
 \end{equation}
This shows, that the unknown parameter $V$ and the 
asymptotically optimal parameter $\eta$ can be directly estimated from the observed spin correlations.

In the next section, we will show that the optimal {\em quadratic} cost function yields in fact 
the {\em total} optimum of the reconstruction error with respect to free variations of the cost function $\Phi$.

\section{The optimal local cost function}
 In this section, we will derive the form of the optimal local cost function $\Phi$ 
 within the cavity/replica approach
 and show that it is
 quadratic. Hence, the results of the previous section can be applied, where the optimal
 quadratic cost function was already computed.
 We will give a derivation of this fact for the case of finite inverse 'temperature' $\nu$,
 assuming that the argument can be continued to $\nu\to\infty$.

 The optimisation of cost functions
 for learning problems within the replica approach
 goes back to the work of Kinouchi and Caticha \cite{kinouchi1992optimal}. We will follow the framework of \cite{advani2016statistical} (see also \cite{Berg_2016}).
 Our goal is to minimise an error measure for a learning problem
 which is of the form $\varepsilon(R,q,q_0)$ such as
 (\ref{eq:error_def_formula}). It depends on order parameters which are computed by
 setting the derivatives of a free energy function $F_{\Phi}(R,q,q_0)$ (such as \ref{eq:free_before_saddle}) equal to zero.
 The main idea is to take these conditions into account within a  Lagrange function
 \begin{equation}
 \varepsilon(R,q,q_0) + \sum_{S \in R,q,q_0} \lambda_S \frac{\partial}{\partial S} F_{\Phi}(R,q,q_0),
 \end{equation}
 where the $\lambda_S$ are the corresponding Lagrange multipliers. The optimal function $\Phi$ is
 obtained from the variation
 \begin{equation}
 \frac{\delta} {\delta \Phi} \sum_{S \in R,q,q_0} \lambda_S \frac{\partial}{\partial S} F_{\Phi}(R,q,q_0) = 0.
 \end{equation}
 For our problem, we can write (see \ref{eq:FG}, \ref{eq:free_before_saddle})
 \begin{equation}
 F_{\Phi}(R,q,q_0) =   F_0(R,q,q_0) - \frac{\alpha}{\nu}
 \int G_{\beta R,q} (v) \ln \Psi_{q_0 - q} (v) dv,
 \end{equation}
 where $F_0(R,q,q_0)$ is independent of $\Phi$ and $G_{\mu, \omega} (v)$ denotes a
 scalar Gaussian density
 with mean $\mu$ and standard deviation $\omega$. The free energy depends on $\Phi$
 through the function
   \begin{equation}
 \Psi_{q_0 - q} (v)  \doteq  \int G_{v, q_0 - q}(y)\; e^{-\nu \Phi(y)} dy.
 \label{relate_Psi_Phi}
 \end{equation}
 We will first derive a condition on the form of the optimal function $\Psi$ from the variation
 \begin{equation}
 \frac{\delta} {\delta \Psi} \sum_{S \in R,q,q_0} \lambda_S \frac{\partial}{\partial S}
  \int G_{\beta R,q} (v) \Psi_{q_0 - q} (v) dv
  = 0.
  \label{free_variation}
 \end{equation}
 From this, we will recover the form of the optimal $\Phi$. To obtain the derivatives with respect to the
 order parameters we use the following rules
 for expectations over Gaussian measures, which can be easily derived
 using integration by parts
 \begin{eqnarray}
\frac{\partial}{\partial \mu} \int G_{\mu, \omega} (v) f(v) dv &= \int G_{\mu, \omega} (v) \partial_v f(v) dv, \\
\frac{\partial}{\partial \omega} \int G_{\mu, \omega} (v) f(v) dv &= \frac{1}{2}
\frac{\partial^2}{\partial \mu^2} \int G_{\mu, \omega} (v) f(v)  \\
&= \frac{1}{2}
\int G_{\mu, \omega} (v) \partial^2 _vf(v) dv.
 \end{eqnarray}
Hence,  the derivatives required for (\ref{free_variation}) are
  \begin{eqnarray}
\frac{d}{dR}  \int G_{\beta R,q} (v) \ln \Psi_{q_0 - q} (v)  dv = \beta  \int G_{\beta R,q} (v) \;
\partial_v \ln \Psi_{q_0 - q} (v) dv, \\
\frac{d}{dq_0} \int G_{\beta R,q} (v) \; \ln \Psi_{q_0 - q} (v) =  \frac{1}{2}
\int G_{\beta R,q} (v) \; \frac{\partial^2_v \Psi_{q_0 - q} (v)}{\Psi_{q_0 - q} (v)} dv  = \\
\frac{1}{2} \int G_{\beta R,q} (v) \; \left\{\partial^2_v \ln \Psi_{q_0 - q} (v) +
\left(\partial_v \ln \Psi_{q_0 - q} (v)\right)^2\right\} dv,
\nonumber\\
\frac{d}{dq}  \int G_{\beta R,q} (v)  \;  \ln \Psi_{q_0 - q} (v) dv =
\frac{\beta^2}{2}  \int G_{\beta R,q} (v)   \;  \partial^2_v  \;  \ln \Psi_{q_0 - q} (v)  dv- \\
 \frac{d}{dq_0} \int G_{\beta R,q} (v) \; \ln \Psi_{q_0 - q} (v) dv.
 \nonumber
\end{eqnarray}
An application of standard variational calculus to a linear combination
of these order parameter derivatives shows that
\begin{equation}
\partial_v \ln \Psi_{q_0 - q} (v) = c_1 + c_2 \partial_v \ln G_{\beta R,q}(v),
\end{equation}
where $c_{1,2}$ are independent of $v$.
Since the logarithm of the Gaussian density $\ln G_{\beta R,q}(v)$ is a quadratic function in $v$, we conclude that also $\ln \Psi_{q_0 - q} (v)$ is a quadratic expression in the variable $v$, making
$\Psi_{q_0 - q} (v)$ a (non--normalised) Gaussian density.

To conclude our argument on the optimal form of $\Phi$, we use
relation (\ref{relate_Psi_Phi}). This shows that the Gaussian density
$\Psi_{q_0 - q} (v)$ is the convolution of a (non--normalised) Gibbs density $e^{-\nu\Phi(y)}$
of a random variable $y$ with the density  $G_{v, q_0 - q}(y) = G_{y, q_0 - q}(v)$ of a Gaussian
random variable $v$. As a convolution corresponds to the addition two random variables, we know that
$v + y$ is also a Gaussian random variable.
Since $v$ is Gaussian, then $e^{-\nu\Phi(y)}$ is also a Gaussian density and $\Phi(y)$
is quadratic in $y$. We have a already computed the best quadratic cost function in the previous Section, and we conclude that
the estimator (\ref{eq:lin_Est}) with (\ref{eq:eta_opt}) is the best local estimator of the couplings.

\section{Cavity approach II: TAP equations and approximate mean field ML estimator}
So far we have ignored the symmetry of the coupling matrix by restricting ourselves
to estimators derived from local cost functions. 
In this Section, we will discuss a well known approximation \cite{Opper2001} of the (symmetric) 
maximum likelihood 
estimator which is based on mean field theory. We will re--derive this estimator using the more advanced
(adaptive) TAP mean field theory, because its results for the spin correlation matrix will also be needed
in the following. We will later
compute its reconstruction error in Section \ref{sec:MF--ML}.
Our starting point is a generalisation
of the well known TAP mean field approach developed for the SK model.
Using the cavity approach \cite{Opper2001}  one  
derives the following 'adaptive' TAP equations for the magnetisations
\begin{equation}
m_i = \tanh\left(\beta\sum_j J_{ij} m_j - \beta^2V_i m_i + \beta H_i\right),
\label{adapt_MF--ML}
\end{equation}
where 
\begin{equation}
V_i = \left\langle \left\{\sum_j J_{ij} (\sigma_j - \langle\sigma_j\rangle_{\backslash i} ) \right\}^2\right\rangle_{\backslash i}
\end{equation}
is the variance of the cavity field at spin $i$. Using a linear response argument 
(i.e. by taking the derivative of $m_i$ eq. (\ref{adapt_MF--ML}) with respect to $H_j$), one obtains the following cavity approximation 
to the susceptibility $\chi_{ij} = C_{ij} - m_i m_j$, i.e. the covariance matrix of the spins:
\begin{equation}
\bm{\chi(J)} = \left(\bm{\Lambda} - \beta \bm{J}\right)^{-1},
\label{chiJ}
\end{equation}
where the diagonal matrix $\bm{\Lambda}$ has elements 
\begin{equation}
\Lambda_{ii} = \beta^2 V_i + \frac{1}{\chi_{ii}} = \beta^2 V_i + \frac{1}{1 - m_i^2}.
\label{Lamdef}
\end{equation}
From this result, we can draw the following conclusions:
\begin{enumerate}
\item Writing the moment matching conditions for the maximum likelihood estimator   as
\begin{equation}
C_{ij}(J) \doteq \langle \sigma_i \sigma_j\rangle=  \hat{C}_{ij} \doteq \frac{1}{M} \sum_{k=1}^{M} \sigma_i^k \sigma_j^k
\end{equation} 
and specialising to the paramagnetic case $H_i = m_i =0$, we have $\bm{C}(\bm{J}) = \bm{\chi}(\bm{J})$. Hence, the cavity approximation 
(\ref{chiJ}) yields the mean field (MF) estimator given by \cite{Kappen}
\begin{equation}
J^{MF}_{ij} = - \frac{1}{\beta} \left(\hat{\bm{C}}^{-1}(\bm{J})\right)_{ij}\quad \textrm{for} \; i \neq j \ .
\label{eq:J_MF--ML}
\end{equation}
At first glance, this simple and explicit form of a (symmetric) coupling estimator does not seem to fit into the 
framework developed in this paper. Surprisingly, we will 
derive a local cost function in the next Chapter which allows for the computation of the reconstruction error
using the statistical physics approach.
\item Inverting  (\ref{chiJ}) and using (\ref{Lamdef}) for $m_i = 0$, we get an expression for the trace of the inverse spin correlation
matrix in terms of the variances of the cavity fields at all spins which is given by
\begin{equation}
\frac{1}{N}\mbox{Tr} \overline{ \bm{C}^{-1}} = \frac{\beta^2}{N} \sum_i V_i + 1 \ .
\end{equation}
If we assume that the teacher couplings $\bm{J}^*$ can be viewed as 
generated from a random matrix 
ensemble for which the $V_i$ become self--averaging, i.e. $V_i \equiv V$ as $N\to\infty$ we finally obtain
the simple result
\begin{equation}
\lim_{N\to\infty}\frac{1}{N}\mbox{Tr} \overline{ \bm{C}^{-1}} =  \beta^2 V + 1 \ .
\label{eq:TRC_Hopfield}
\end{equation}
With this result, we can eliminate another unknown parameter of the teacher's ensemble of couplings,
as we have shown that $V$ can be estimated from the observed  spin data, see  (\ref{eq:Delta}) and
(\ref{delta_explicit}).

Equation (\ref{eq:TRC_Hopfield}) agrees with the special result (\ref{SK_inverse}) for the SK model, since the 'Onsager correction' in the TAP equations for the SK model
gives $V=1$.
As an application of the general result  (\ref{eq:TRC_Hopfield}), we present numerical results for the reconstruction error
for the  Wishart ensemble in Section \ref{sec:results}, where the couplings are given by
\begin{equation}
J^*_{ij}  = \frac{1}{ N} \sum_{\mu=1}^{\gamma N} \xi_i^\mu \xi_j^\mu
\label{eq_HW}
\end{equation}
and the $\xi_j^\mu$ are independent zero mean Gaussian random variables with unit variance.
The thermodynamics of this model agrees with that of the  celebrated Hopfield model of a neural network 
(where $\xi_i^\mu = \pm 1$)   \cite{hopfield}, in the phase where there is no macroscopic overlap
between the spin configurations and a stored pattern.  Hence, we can read off the cavity variance
from the TAP mean field equations obtained by \cite{Seung-Sompolinsky-Tishby}, setting $m_i=0$. One finds
\begin{equation}
V  =  \frac{\gamma}{1 - \beta} .
\label{eq:V_Hopfield}
\end{equation}
\end{enumerate}
For other random matrix ensembles which are invariant against orthogonal transformations
it is possible to obtain a general expression for the cavity variance in terms of the so--called R--transform of the matrix ensemble (for details, see \cite{parisi1995mean,opper2016theory}) and can be expressed by the
limiting eigenvalue spectrum of the matrices.

\section{Reconstruction  error for MF--ML estimator}
\label{sec:MF--ML}
We will now turn to the computation of the reconstruction error for the MF--ML estimator
(\ref{eq:J_MF--ML}).
At first glance, this estimator does not seem to be related to a local cost function 
in the style of (\ref{eq:en_fz}). But surprisingly, it is not hard to construct such a function.
If we specialise again to the estimation of the coupling vector
$\bm{W}$ corresponding to spin $\sigma_0$, we can simplify the estimator (\ref{eq:J_MF--ML}) using 
the matrix inversion lemma \cite{matrix_book}  in the form
\begin{equation}
W^{MF}_{i} = \sqrt{N} J^{MF}_{0i} =  - \frac{\sqrt{N}}{\beta}(\hat{C}^{-1})_{0i} =   
\frac{\sqrt{N}}{\beta} \phi_0 \sum_{j\neq 0} (\hat{C}_{-0}^{-1})_{ji} \hat{C}_{0j} \qquad i\neq 0,
\label{eq:MF_estimator}
\end{equation}
where
\begin{equation}
\phi_0 = \frac{1}{1 -  \sum_{i,j\neq 0} \hat{C}_{0i}(\hat{C}_{-0}^{-1})_{ij} \hat{C}_{0j}  } = \frac{1}{1 - \Delta},
\label{def_phi}
\end{equation}
where $\Delta$ was introduced in (\ref{eq:Delta}). Assuming as before, that $\Delta$ is self--averaging
for $N\to\infty$, the mean field estimator is of the form (\ref{eq:lin_Est}) and is
associated to a cost function of the form (\ref{eq:cost_fz_quadratic}). Hence, the results of sections \ref{sec:linear} apply. In particular, from (\ref{delta_explicit}) and (\ref{def_phi}) we compute
\begin{equation}
\phi_0 = \frac{1}{1 -\Delta} =  \frac{\alpha (1+\beta^2 V)}{\alpha - 1},
\end{equation}
and the estimation error is given by (\ref{eq:error_quad}) with the parameter
$
\eta_{MF} = \phi_0/\beta 
$:
\begin{equation}
\varepsilon_{MF}= \frac{Y}{(\alpha -1)^2}+\frac{ \alpha^2}{\beta^2 (\alpha-1)^3} (1+\beta^2 V) \frac{1}{N} Tr
\overline{\bm{C}^{-1}}.
\label{eq:error_MF--ML}
\end{equation}

\section{Asymptotics}
\label{sec:asymptotics}
We will now investigate the limiting scaling of the reconstruction error as the number of data $M$
grows much larger than the number of parameters (per spin) $N$ to be estimated. This means 
we consider the limit $\alpha\to\infty$. This is of special interest, because we can compare the
results obtained by our replica/cavity approach with
results derived independently by standard arguments of classical statistics. 
From (\ref{eq:err_opt}), (\ref{eq:error_MF--ML})
and Appendix  \ref{sec:AppD} we can see that, as $\alpha\to\infty$, the scaling of the reconstruction errors 
for the pseudo--likelihood estimator, the optimal local estimator and the mean field estimator is
\begin{equation}
\varepsilon \simeq \frac{c}{\alpha},
\label{eq:as_err}
\end{equation}
where
\begin{eqnarray}
c_{\mbox{\tiny{PLM}}}  &= &
\frac{1}{ \beta^2 } \frac{1}{\int d v \; G_{\beta V,V} (v) \left( 1- \tanh^2(\beta v) \right) }
\; \frac{1}{N} \mbox{Tr}  \overline{C^{-1}},
\label{asympseudo-}\\
c_{\mbox{\tiny{OPT}}} & = & \frac{1 + \beta^2 V}{\beta^2} \frac{1}{N} \mbox{Tr} \overline{ \bm{C}^{-1}} =
   \frac{(1 + \beta^2 V)^2}{\beta^2}, \label{asymp--OPT} \\
c_{\mbox{\tiny{MF--ML}}}  & = & c_{\mbox{\tiny{OPT}}},
\label{asympMF--ML}
\end{eqnarray}
where in the second equality of the second line, we have used (\ref{eq:TRC_Hopfield}).  
Hence, asymptotically the
simple mean field estimator and the optimal estimator converge to the true couplings at the same 
speed. Thus, one might conjecture that the 
mean field estimator is equivalent to the true maximum likelihood estimator in the thermodynamic limit,
assuming that the cavity approach is correct.

The validity of the inequality
$c_{\mbox{\tiny{PLM}}} > c_{\mbox{\tiny{OPT}}}$ given by  (\ref{asympseudo-}) and (\ref{asymp--OPT})
will depend on the temperature $\beta$ and can be established at least for $\beta$ small enough.
 For the SK model, this region covers the entire 
paramagnetic phase $\beta < 1$ where our simple cavity method is valid. 
However, the difference between the two is not very large for small $\beta$. In fact,
expanding (\ref{asympseudo-}) in powers of $\beta$ shows that error coefficients $c$ for both estimators
agree up to terms of order $\beta ^2$. 
One may however argue that the comparison between the two
estimators is not fair, because the pseudo--likelihood estimator does not yield symmetric couplings $J_{ij}$ 
whereas the mean field one (and hence, asymptotically the optimal one) does.
One might thus get a better estimate by a final symmetrisation. Unfortunately, with our present method, the effect of    
symmetrisation on the reconstruction error cannot be computed. We expect that methods of random matrix 
theory would be needed for this. Hence, we postpone a treatment of this problem to future publications.
On the other hand, preliminary simulations show that the improvement of the pseudo likelihood estimator
after symmetrisation is rather weak (at least for the systems with random couplings studied in this paper). This result is further supported by the fact that for small $\beta$, the pseudo likelihood estimator is already {\em almost symmetric}, a fact that can be easily shown, if we expand (\ref{pseudo-_cost}) for small $\beta$. The lowest order term yields an explicit result which is symmetric.

We want to compare the replica based asymptotics (\ref{asympseudo-}) and (\ref{asymp--OPT}-\ref{asympMF--ML}) 
with exact asymptotic expressions for the errors
of statistical estimators which are defined by the minimisation of smooth cost functions of the type
(\ref{eq:en_fz}), see e.g. \cite{schervish2012theory} or \cite{Seung-Sompolinsky-Tishby} for an alternative derivation using replicas.
The idea behind such asymptotic results is an  expansion of the cost function 
in terms of the parameters $\bm{W}$ around the teacher parameters $\bm{W}^ *$ (assuming 
convergence to the teacher in the infinite data limit).
Setting $\delta \bm{W} \doteq \bm{W} - \bm{W}^ *$ and using the law of large numbers and central limit 
arguments one can show the following equation for the data averaged correlations 
\begin{equation}
\overline{\delta W_i  \delta W_j}\simeq \frac{1}{N \alpha} [ (\bm{U}^{-1} \bm{B} \bm{U}^{-1})_{ij}]\quad \mbox{for}\; \alpha\to\infty,
\label{eq:as_err2}
\end{equation}
together with $\overline{\delta \bm{W}} \simeq 0$. The matrices are given by
\begin{equation}
\begin{split}
U_{ij} & = \left\langle\partial_i \partial_j \mathcal{E}(\bm{W}^*; \bm{\sigma}) \right\rangle, \\
B_{ij} & =   \left\langle\partial_i   \mathcal{E}(\bm{W}^*; \bm{\sigma}) \partial_j \mathcal{E}(\bm{W}^*; \bm{\sigma})
\right\rangle.
\label{eq:V_matrix}
\end{split}
\end{equation}
The partial derivatives are with respect to the components of $\bm{W}^*$ and the brackets 
are averages over spins using the distribution $P(\bm{\sigma} | \bm{J}^*)$. For the pseudo--likelihood case,
(\ref{eq:as_err2}) can be further simplified.  In Appendix \ref{sec:AppE}, we show that in this case $\bm{U}\equiv-  \bm{B}$ and we 
finally obtain 
\begin{equation}
\varepsilon  \simeq N^{-1} \sum_i \overline{   (\delta W_i)^2 }= \frac{1}{N \alpha} \mbox{Tr}B^{-1},
\label{eq:final_error}
\end{equation}
with
\begin{equation}
B_{ij} = \beta^2   \left\langle  \sigma_i \sigma_j \left[1-\tanh^2 (\beta \frac{1}{\sqrt{N}} \sum_{j \neq 0}W^*_{0j} \sigma_j ) 
\right] \right\rangle.
\end{equation}
If we neglect the correlations between $\sigma_i \sigma_j $ and the field $\frac{1}{\sqrt{N}} \sum_{j \neq 0}W^*_{j0} \sigma_j $ for large $N$ and note that $\langle  \sigma_i \sigma_j \rangle = C_{ij} $, this result is in agreement with (\ref{asympseudo-}).

A similar calculation is possible for the OPT/MF--ML case. Here we get 
\begin{equation}
\begin{split}
U_{ij} & =  \frac{\beta}{\phi_0 N} C_{ij},\\
B_{ij} & =   \frac{\beta^2}{\phi_0^2 N} \langle \sigma_i \sigma_j h_*^2 \rangle + \frac{1}{N} C_{ij} 
- 2 \frac{\beta}{\phi_0 N} \langle h_* \sigma_0 \sigma_i \sigma_j \rangle.
\end{split}
\end{equation}
To obtain the asymptotics of the replica result (\ref{asymp--OPT}) form these
matrices,
we assume that the dependencies between the random variables $\sigma_i \sigma_j$ on the one hand
and  
respectively  $h_*^2$ and $h_* \sigma_0$ on the other hand can be neglected for $N\to\infty$.  Using the facts that $\langle h_*^2\rangle = \beta^2 V^2 + V$,
$\langle h_* \sigma_0\rangle = \beta V$ and $\lim_{\alpha\to\infty} \phi_0 = 1 + \beta^2 V$
finally yields (\ref{asymp--OPT}) .

\section{Numerical Results} \label{sec:results}
In the previous Sections we saw that the error of any algorithm that infers the network couplings by minimizing a cost function of the kind  (\ref{eq:def_Phi}) satisfies (\ref{eq:error_def_formula}), in the large $N$ limit, when 
the cavity arguments apply. The order parameters are the ones extremizing the free energy (\ref{eq:free_energy}). 
For pseudo--likelihood maximization, the set of equations (\ref{eq:y_sp}) for the order parameters has to be solved numerically, whereas for the local optimal and MF--ML estimators we computed analytically the error in the form, respectively, (\ref{eq:err_opt}) and (\ref{eq:error_MF--ML}).
Note that the  error (\ref{eq:error_def_formula}) is expressed in terms of three parameters that depend on the distribution of the teacher couplings: $Y$, $V$ and the trace of the inverse correlation matrix $\bm{C}$. As an example, we considered the the Gaussian ensemble of the SK model, with parameters given by $Y=1$ and  relation (\ref{SK_inverse}) and the Wishart ensamble of (\ref{eq_HW}) with parameters given by $Y=\gamma$  and  relations  (\ref{eq:TRC_Hopfield}) and (\ref{eq:V_Hopfield}). 
Figure \ref{fig:error} compares the predicted error with the mean squared error that we get from simulations, as a function of $\alpha$. We show results for the pseudo--likelihood, local optimal and MF--ML algorithms applied to the SK and  the Wishart model.
We only report results for the high--temperature (paramagnetic) region, i.e. for $\beta<\beta_c$ where $\beta_c$ defines the onset of spin--glass
ordering. In this region, we expect that on the one hand, the cavity arguments are exact and the other hand,
the convergence of the spin simulations to the thermal equilibrium is sufficiently fast. For the SK model, we have 
$\beta_c =1$ and for the Wishart model $\beta_c \simeq 1/(1+\sqrt{\gamma})$ for zero magnetization and small $q$  \cite{amit_finitep}, where $q$ is the Edwards-Anderson order parameter.
The data are generated by Monte Carlo sampling with a burn in time of $10^7 N$ spin updates and sampling every $10 \, N$ updates, and the couplings are recovered either by minimizing the pseudo--likelihood cost function (\ref{pseudo-_cost}) using a Newton method or from the empirical correlation matrices: see (\ref{eq:lin_Est}, \ref{eq:eta_opt}) for the local optimal algorithm and (\ref{eq:J_MF--ML}) for MF--ML.
The plot shows that the replica calculation predicts rather well the results of the simulations for systems of $N=100$ spins.
In addiction, it is clear that the optimal local algorithm outperforms the other two methods and,
 in the high--temperature regime considered here, the MF--ML algorithm performs better than pseudo--likelihood maximization. This performance difference is   more relevant for increasing  $\beta$ and in the small $\alpha$ region, whereas it is almost negligible for large $\alpha$, in agreement with the asymptotic expansions. Finally,  we compare the analytical results for 
the asymptotic behavior of the error computed in Section \ref{sec:asymptotics} with the results from simulations. 
Assuming the scaling  (\ref{eq:as_err}),  
we fitted the function $\varepsilon=c/\alpha$ to the mean squared  error of the  couplings inferred from simulations at large $\alpha$.  In Table \ref{table1} we show that this 'experimental' value of $c$ is consistent with $c_{\mbox{\tiny{PLM}}} $ (\ref{asympseudo-}) and $c_{\mbox{\tiny{OPT}}} =c_{\mbox{\tiny{MF--ML}}}$ (\ref{asymp--OPT}, \ref{asympMF--ML}).
We then plot the predicted value of $c$ as a function of $\beta$ in 
 Figure \ref{fig:Asympt}, where we can see that the difference between pseudo--likelihood maximization, the local optimal and  MF--ML algorithms is almost indistinguishable  and goes to zero for small $\beta$, as we would expect by noticing that the analytical formula for $c_{\mbox{\tiny{PLM}}} $ (\ref{asympseudo-}) agrees with $c_{\mbox{\tiny{OPT}}} =c_{\mbox{\tiny{MF--ML}}}$ (\ref{asymp--OPT}, \ref{asympMF--ML}) up to second order in $\beta$.
 From the plot it is also clear that for larger $\beta$ --i.e. smaller stochasticity of the spins-- the error in predicting the couplings is smaller.

The three algorithms show different behaviors in the small $\alpha$ region. 
As the MF--ML algorithm relies on the inversion of the correlation matrix (\ref{eq:J_MF--ML}), that becomes singular at $\alpha=1$, its error diverges at $\alpha=1$, as can be seen from (\ref{eq:error_MF--ML}). On the contrary, the error of the optimal local algorithm shows no divergence, 
since $\eta_{\mbox{\tiny{opt}}}=0$ and $\varepsilon=Y$ at $\alpha=1$ (see \ref{eq:eta_opt}, \ref{eq:err_opt}).
From simulations we also observe that the error of the pseudo--likelihood estimator increases for decreasing $\alpha$ and for $\alpha<2$ it reaches large values, with large variations across trials, while the extremization of the free energy (see \ref{eq:y_sp}) fails in the region $\alpha < 2$.
A way to overcome this divergence is to introduce a regularizing term in the objective function. We postpone the study of regularized estimators to future work.
We present additional plots in Appendix \ref{sec:AppF}, showing the error dependence on the system size.

\begin{figure}[htb]

\centering
\vspace*{1\baselineskip}
\includegraphics[width=0.48\linewidth]{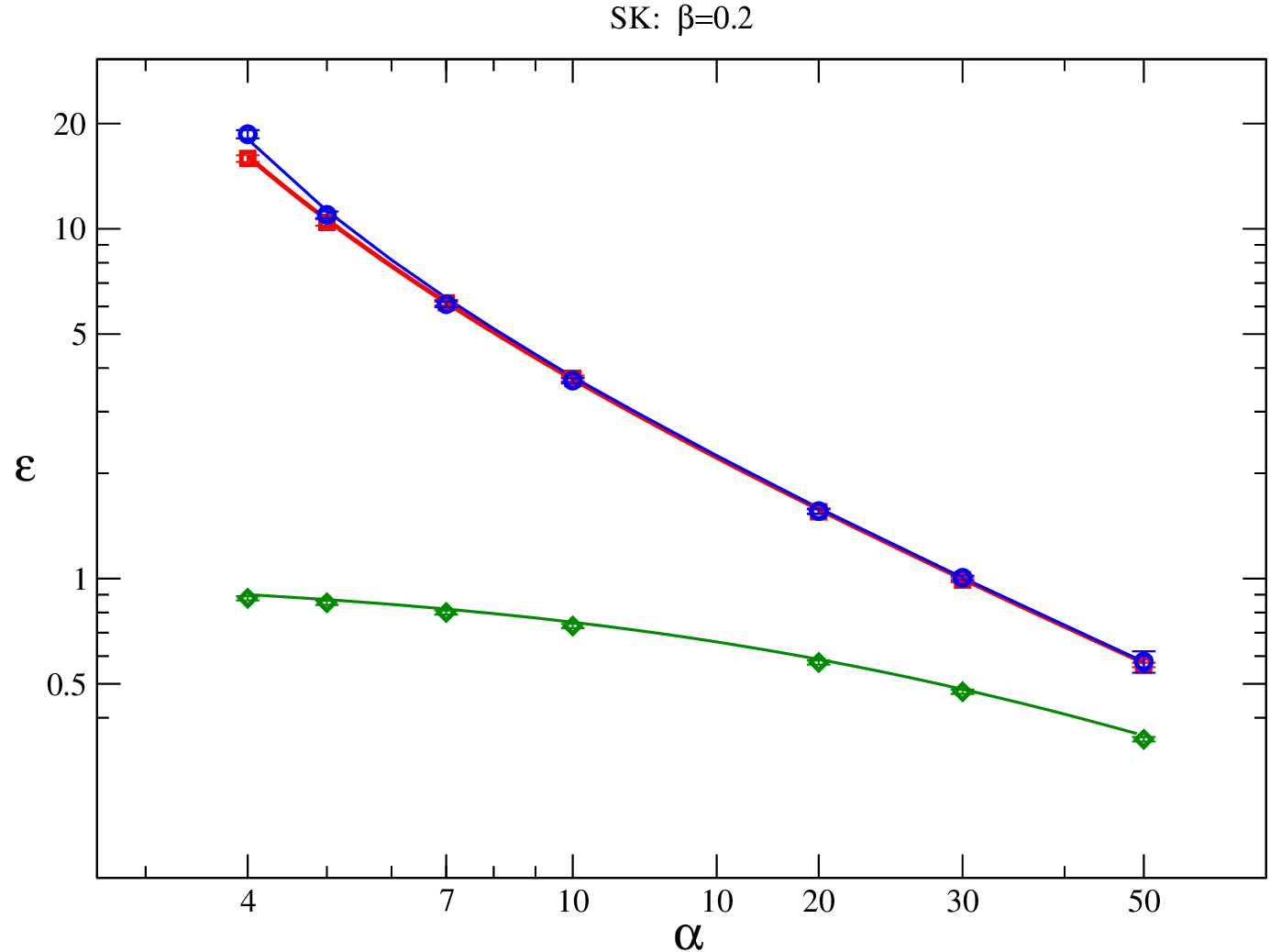} \hfill
\includegraphics[width=0.48\linewidth]{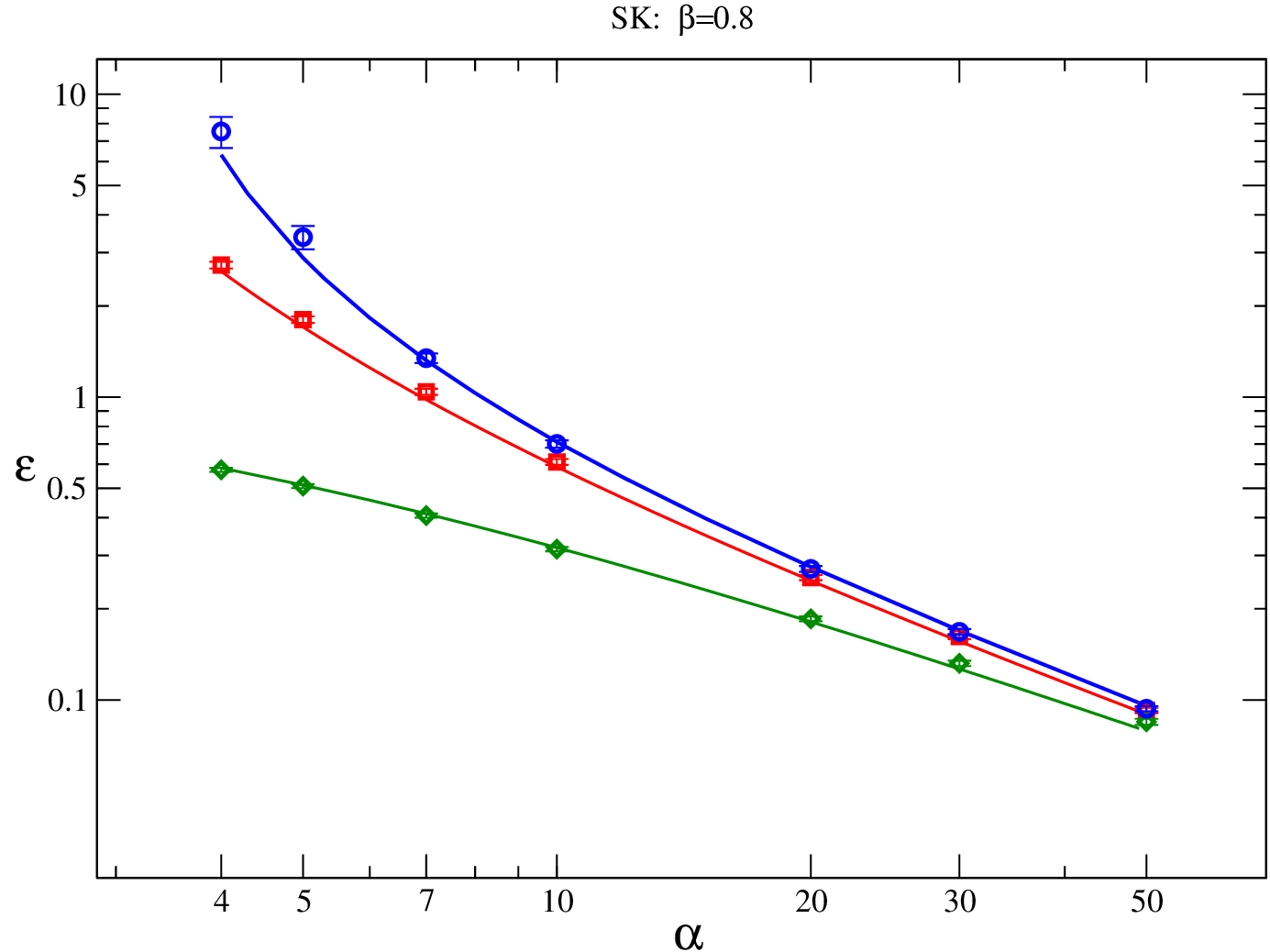}\\
\vspace*{2\baselineskip}
\includegraphics[width=0.48\linewidth]{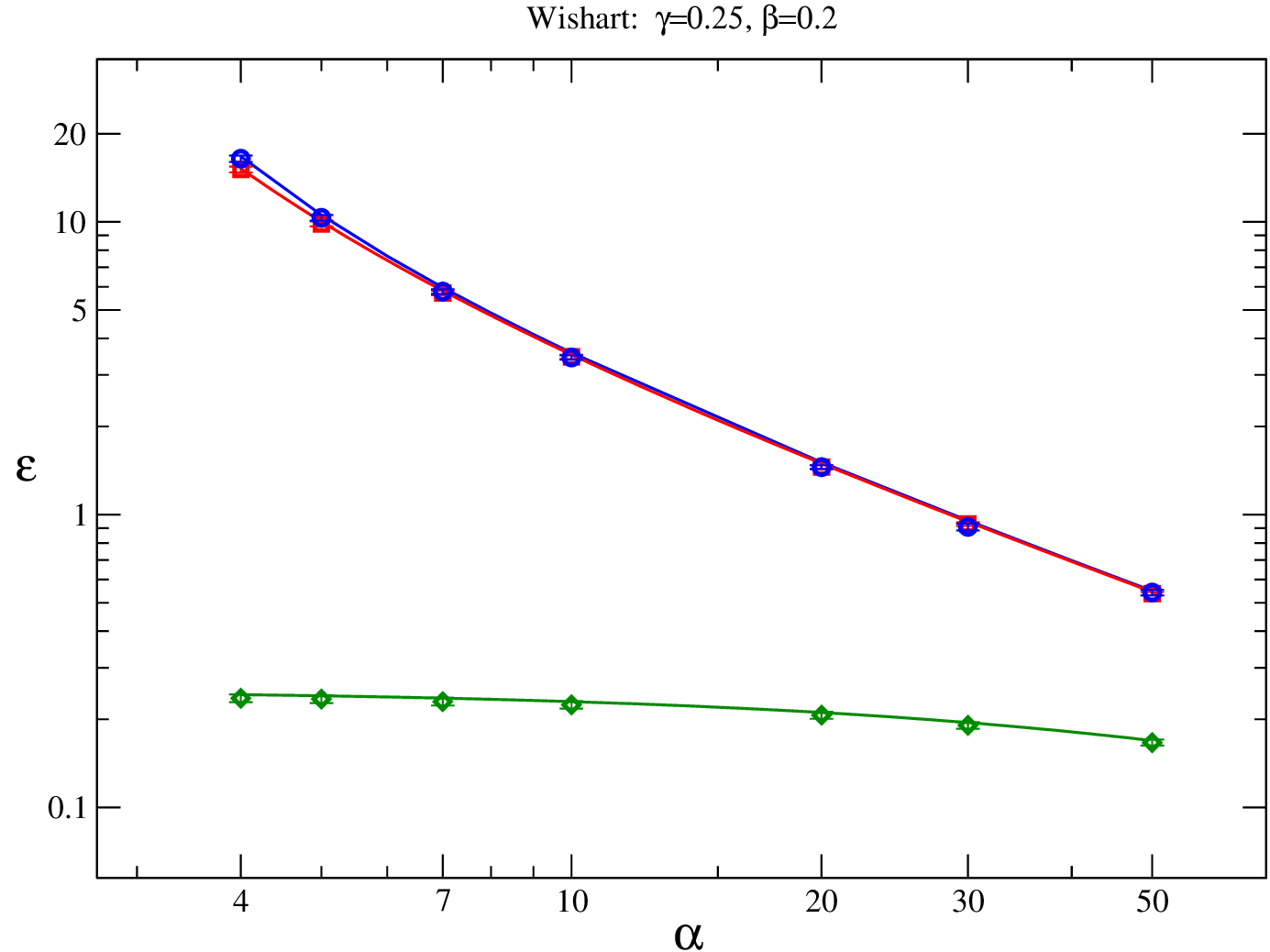} \hfill
\includegraphics[width=0.48\linewidth]{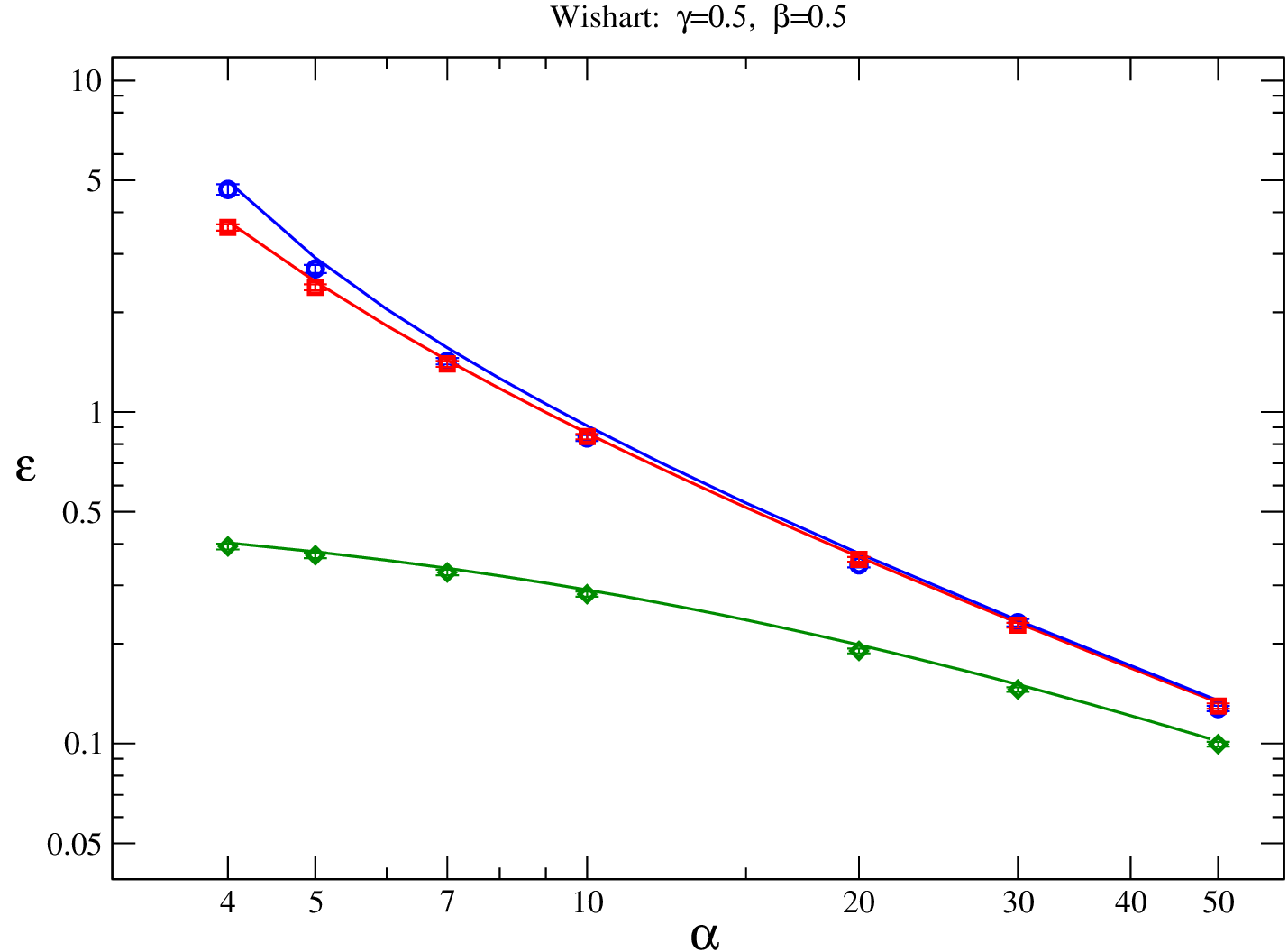}

\caption{The mean squared error of the couplings inferred by using the pseudo--likelihood algorithm (blue dots), the optimal local algorithm (green dots)  and the MF--ML algorithm (red dots) is compared to the corresponding average prediction error from the replica calculation (continuous lines). The error is plotted as a function of $\alpha$. Four different systems are considered: SK model at $\beta=0.2$ (top left), SK model at $\beta=0.8$ (top right), Wishart model with $\gamma=0.25$ at $\beta=0.2$ (bottom left) and Wishart model with $\gamma=0.5$ at $\beta=0.5$ (bottom right). The algorithms were tested on a system of $N=100$ spins and the results are averaged over $5$ realizations of the network and $100$ different datasets generated from each network.  Error bars represent standard deviations of the means. }
\label{fig:error}
\end{figure}

\begin{figure}[htb]

\centering
\vspace*{2\baselineskip}
\includegraphics[width=0.6\linewidth]{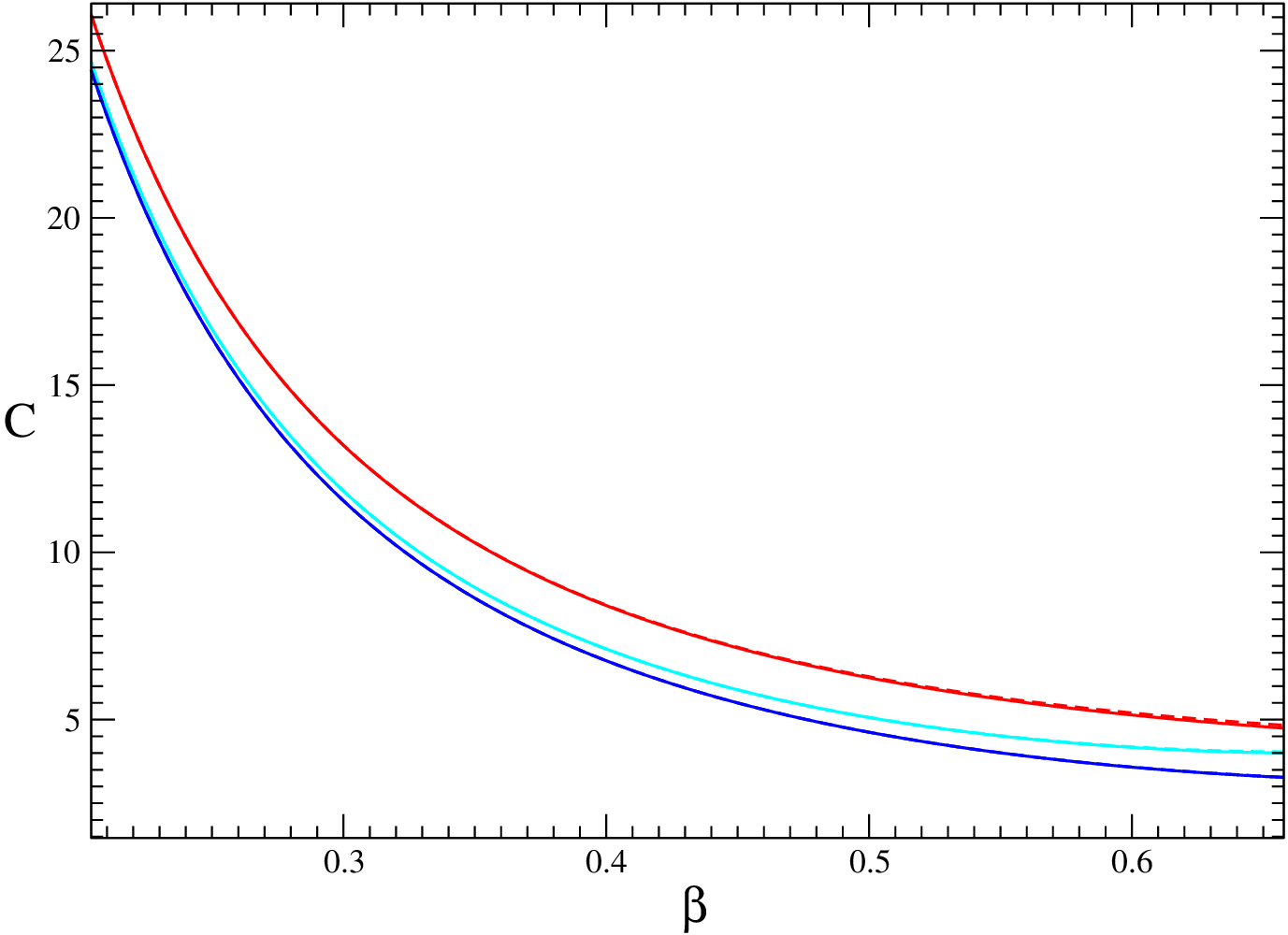}\\

\caption{The values 
$c_{\mbox{\tiny{PLM}}} $ (\ref{asympseudo-}) for pseudo--likelihood maximization (dotted lines) and  $c_{\mbox{\tiny{MF--ML}}} $ (\ref{asympMF--ML}) for the MF--ML algorithm (continuous lines)   are plotted as a function of $\beta$. The red lines correspond to the SK model, the  blue lines to the Hopfield model with parameter $\gamma=0.25$ (light blue) and $\gamma=0.15$ (dark blue).
}
\label{fig:Asympt}
\end{figure}

\begin{table}[h!]
\begin{tabular}{|c|c|c|c|}
\hline 
Model & Algorithm & c  & c (simulations) \tabularnewline
\hline 
\hline 
SK & PLM &  $5.199$ & $ 5.16 \pm 0.04 $ 
\tabularnewline
\hline 
 & MF--ML &  $5.137$ & $ 5.14 \pm 0.05$ 
\tabularnewline
\hline 
Wishart & PLM  &  $3.582$ & $ 3.64 \pm 0.06$
\tabularnewline
\hline 
 & MF--ML &  $3.578$ & $ 3.60 \pm 0.05$ 
\tabularnewline
\hline 
\end{tabular}
\caption{
The values of $c_{\mbox{\tiny{PLM}}} $ (\ref{asympseudo-}) for pseudo--likelihood maximization and  $c_{\mbox{\tiny{MF--ML}}} $ (\ref{asympMF--ML}) for the MF--ML algorithm  are compared to the results 'c (simulations)' we obtained by fitting  the function $\varepsilon=c/\alpha$ to  the  mean squared  error $\varepsilon$ of the inferred couplings   obtained from simulations at large $\alpha$. We considered two systems: SK model at $\beta=0.6$ and Hopfield model with $\gamma=0.15$ at $\beta=0.6$. The algorithms were tested on a system of $N=200$ spins for $\alpha=900, 950, 1000$ and the results are averaged over $5$ realizations of the network and $10$ different datasets generated from each network. 
}
\label{table1}
\end{table}

\section{Discussion and outlook}
We have presented a statistical physics approach for calculating the reconstruction error of 
algorithms for learning the couplings of large Ising models. Our method 
assumes local cost functions for learning and is based on  a combination of
the replica trick and of cavity arguments for computing quenched averages over spin configurations
which are drawn at random from a teacher network. A replica symmetric ansatz seems to be justified  as long as the learning algorithms are based on convex cost functions.
The cavity approach assumes 
a large densely connected network with couplings that are roughly of the 
same size leading to only weakly correlated spins. These assumptions are correct in the thermodynamic 
limit for certain statistical ensembles of network couplings but may also give good approximations for realistic networks. 
While our method is so far restricted to problems 
which are realisable by pairwise spin--interactions, it could nevertheless be of practical interest in providing approximate statistics for hypotheses testing against more complicated network models
 (having e.g. 3--spin interactions).

Our results show that the learning problem is, at least within our framework, surprisingly simple:
An explicit estimator based on a quadratic cost function achieves minimal error and 
outperforms the more complicated  pseudo--likelihood estimator. This 
optimal estimator only requires prior knowledge of the length of the true coupling vector.
Moreover, a simple (symmetric) mean field approximation to the maximum likelihood estimator is asymptotically
optimal and can be computed without such prior knowledge.
In the case of the SK model, the region of small $\beta$ in which these results 
hold covers the entire paramagnetic
phase, where our simple cavity arguments are known to be valid. It would be interesting
to work out analytically how well the mean field estimator approximates the exact maximum likelihood
estimator in the thermodynamic limit.

Our work  is only a  first step to an understanding of the typical performance of learning algorithms for the inverse
Ising problem. From a technical point of view our method could be generalised in several directions.
We have restricted ourselves to models where data are sampled from the paramagnetic phase of a
teacher network. While it is possible to generalise the analysis, the average over samples from
a spin--glass phase would usually require more complex 
types of cavity arguments \cite{Mezard_Parisi_2001} which are related to the breaking of the replica symmetry
of the teacher network.  In such a case, the simple Gaussian distribution of cavity fields on which our analysis
strongly relies is no longer valid. One might expect that now the quadratic cost functions
may no longer be optimal (and not even consistent) but could be outperformed by a pseudo--likelihood
method.

We also expect that our cavity framework could be extended to 
sparse networks as long as the number of nonzero couplings per spin is large enough
to allow for the application of the central limit arguments used in our work.

After finishing our work we became aware of a recent preprint \cite{Berg_2016} where similar learning 
problems (focussing on a teacher model with independent Gaussian couplings) were studied.
The author applied a double replica calculation (the other set of replica are used for dealing with the
partition function in the quenched average over the spins) instead of using cavity arguments. This  
results in a free energy function which agrees essentially with our result
(\ref{eq:free_energy}). However, the order parameters appearing in the free energy are not  defined by (\ref{eq:def_parameters}) but by (\ref{eq:def_rho}) instead,  
and the reconstruction error differs from ours. The major difference is that the result for the 
error in \cite{Berg_2016} does not contain the spin--correlation matrix as in our equation (\ref{eq:error_def_formula}). We believe that  this could be related to an implicit approximation of the correlation matrix by a unit matrix.

\section*{Acknowledgements}
We would like to thank Andrea Pagnani for fruitful discussions that significantly motivated our work.

\begin{appendices}

\section{Details of the replica calculation}
\label{sec:AppA}
From (\ref{eq:replica_trick}-\ref{eq:Z_n}-\ref{eq:F_inner_average}) one can see that
  the free energy
 can be written as
\begin{equation}
\begin{split}
F= - \lim_{n \to 0}  N^{-1}\nu^{-1} \frac{\partial}{\partial n}  \ln & \int   \prod_{a=1}^n  \; d \bm{W}^a     \left\{
  \sum_{\sigma_0} \int  \,  dh_* \prod_{a=1}^n dh_a       
\frac {1}{Z_0}  \exp \left[ \beta \sigma_0 h_* \right] \right. \\
& \left. \exp \left[- \nu  \sum_{a=1}^n \Phi( \sigma_0 h_a) \right]  \;  
p_{\mbox{\tiny{cav}}}(h_*,h_1,\ldots h_n) \right\}^{\alpha N},
 \end{split}
\label{eq:F_complete}
\end{equation}
where $p_{\mbox{\tiny{cav}}}(h_*,h_1,\ldots h_n)$ is a multivariate Gaussian density with zero mean
and covariance given by (\ref{eq:cav_h}).
The average over the Gaussian fields yields quadratic terms in $\bm{W}^*$ and $\{ \bm{W}^a \}_{a=1}^n$, 
that can be simplified by introducing the order parameters $\{ R,q,q_0 \}$ (\ref{eq:def_parameters}), that have to be defined via integrals over delta functions.
One finds that free energy decouples into two terms:
\begin{equation}
{F}(R,q,q_0)=    F_0(R,q,q_0) +F_1(R,q,q_0).
\label{eq:FG}
\end{equation}
The first one contains the integrals over the couplings and  measures the density of the networks with order parameters $R,q,q_0$ :
\begin{equation}
F_0(R,q,q_0) = -   \lim_{n \to 0}  \nu^{-1} \frac{\partial}{\partial n}  \frac{1}{N}\ln Z_{\tiny{\mbox{coup}}},
\label{eq:G0}
\end{equation}
with
\begin{equation}
\begin{split}
Z_{\tiny{\mbox{coup}}} &= \int  \prod_a d\bm{W^a}
 \prod_{a } \delta\left(\sum_{ij} W^a_i C_{ij} W^*_j - N R\right) \\
& \prod_{a } \delta\left(\sum_{ij} W^a_i C_{ij} W^a_j - N q_0\right)
 \prod_{a < b} \delta\left(\sum_{ij} W^a_i C_{ij} W^b_j - Nq\right). \\
\end{split}
\end{equation}
For notational simplicity  here we have dropped  the '$0$' from the correlation matrix $\bm{C}^{\backslash 0}$.
$F_0$ can be computed following our derivation in \cite{LBR_MO_15}: we introduce the orthogonal matrix $\bm{U}$ that diagonalizes 
$\bm{C} = \bm{U}^\top \Lambda \bm{U}$,
\begin{equation}
\begin{split}
Z_{\tiny{\mbox{coup}}} &= \int   \prod_a d\bm{W^a}
 \prod_{a } \delta(\sum_{ijk} U_{ij}  W^a_j \Lambda_i U_{ik} W^*_k  - N R) \\
& \prod_{a } \delta(\sum_{ijk} U_{ij}  W^a_j \Lambda_i U_{ik} W^a_k  - N q_0)
 \prod_{a < b} \delta(\sum_{ijk} U_{ij}  W^a_j \Lambda_i U_{ik} W^b_k  - Nq), \\
\label{eq:Z_coup}
\end{split}
\end{equation}
and transform the student coupling vector into new variables $\bm{U}^\top \bm{W}^a  \to \bm{W}^a$, which we give the same name. We then  express the delta functions as integrals  over the auxiliary parameters $\{  \hat{R} ,\hat{q},  \hat{q_0} \} $.
The integration gives
\begin{equation}
\begin{split}
F_0(R,q,q_0) =  \underset{\hat{R} ,\hat{q},  \hat{q_0}}{\mbox{extr}} \; \frac{1}{ \nu} & \left\{
i \hat{q_0} q_0 +i \hat{R} R - \frac{i}{2}   \hat{q} q
+ \frac{1}{2N} \sum_i 
\frac{ \hat{q}  + i \hat{R} ^2  (\sum_j U_{ij} W^*_j)^2 \Lambda_i}{ 2 \hat{q_0} - \hat{q}} \right. \\
&\left. + \frac{1}{2N} \sum_i  \ln \left[   \Lambda_i (i\hat{q} - 2 i \hat{q_0}) \right] -  \frac{1}{2} \ln(2 \pi) \right\}
\end{split}
\label{eq:G_0_int_step}
\end{equation}
and the extremum over the conjugate order parameters yields
 \begin{equation}
F_0(R,q,q_0) = \frac{1}{2 \nu} \left[ \frac{q_0-R^2/V}{q-q_0}-   \ln (q_0-q) + \frac{1}{N} Tr \ln \overline{C} \right],
\end{equation}
where $V$, representing the cavity variance of the teacher field $h_*$, was introduced in (\ref{eq:def_parameters}).
The second term of (\ref{eq:FG}) contains the integration over the cavity fields $h_*$ and $\{ h_a \}_{a=1}^n$:
\begin{equation}
\begin{split}
F_1(R,q,q_0)= - \lim_{n \to 0}  N^{-1}\nu^{-1} \frac{\partial}{\partial n}  \ln &    \left\{
 2 \int  \,  dh_* \prod_{a=1}^n dh_a       
\frac {1}{Z_0}  \exp \left[ \beta h_* \right] \right. \\
 \exp \left[- \nu  \sum_{a=1}^n \Phi(  h_a) \right]  \;  
& \left. p_{\mbox{\tiny{cav}}}(h_*,h_1,\ldots h_n)
 \right\}^{\alpha N},
 \end{split}
\end{equation}
where we applied the change of variables $\sigma_0 h_* \to h_*$ and $\sigma_0 h_a \to h_a$. The integration gives
\begin{equation}
\begin{split}
F_1(R,q,q_0) =& - \frac{\alpha}{\nu}  \int \frac{d v}{\sqrt{2 \pi q}}  \; e^{-\frac{(v-\beta R)^2}{2 q}} \;
 \ln \int  \frac{d y}{\sqrt{2 \pi (q_0-q)}} \; e^{-\frac{(y-v)^2}{2 (q_0-q)}} e^{- \nu \Phi(y)}.
 \end{split}
 \label{eq:G1}
\end{equation}
Hence the free energy (\ref{eq:FG}) becomes
\begin{equation}
\begin{split}
{F}(R,q,q_0)= -  &   \frac{1}{\nu}  \left\{ \frac{1}{2}  \frac{q_0-R^2/V}{q_0-q} + \frac{1}{2} \ln(q_0-q)  - \frac{1}{2N} Tr \ln  \overline{C}  \right. \\
& \left. + \alpha  \int d v  \; G_{\beta R,q} (v) \;
 \ln \int   d y \; G_{v,q_0-q} (y) e^{- \nu \Phi(y)}
 \right\},
 \end{split}
 \label{eq:free_before_saddle}
\end{equation}
 where $G_{\mu, \omega} (v)$ denotes a
 scalar Gaussian density
 with mean $\mu$ and standard deviation $\omega$.

\section{Saddle point equations for the order parameters}
\label{sec:AppA2}
We rewrite (\ref{eq:free_energy}) as
\begin{equation}
F= -  \underset{q,R,x}{\mbox{extr}} \,  \left\{ \frac{1}{2}  \frac{q - R^2/V}{x}  
 +   \alpha  \int \mathcal{D} v   \;\max_y \left[ -\frac{(y-\sqrt{q} v-\beta R)^2}{2 x} - \Phi(y) \right] \right\},
\end{equation}
where $\mathcal{D} v= e^{-v^2/2}/\sqrt{2 \pi}$. 
The extremum over the order parameters gives the following set of equations:
\begin{equation}
\begin{split}
&0 = \frac{1}{x}- \frac{\alpha }{\sqrt{q}} \int \mathcal{D} v \; v \;   \left. \frac{\partial  \Phi(y)}{\partial y}\right\vert_{y=\hat{y}} \\
&0 = -\frac{R}{V x}- \alpha \beta \int \mathcal{D} v \; \left. \frac{\partial   \Phi(y)}{\partial y}\right\vert_{y=\hat{y}} \\
 &0 = -\frac{1}{x^2} \left( q-\frac{R^2}{V} \right) + \alpha \int \mathcal{D} v \; \left(\left. \frac{\partial \Phi(y)}{\partial y}\right\vert_{y=\hat{y}} \right)^2,
 \end{split}
\end{equation}
where 
\begin{equation}
\hat{y}= \mbox{arg} \max_y \left[ -\frac{(y-\sqrt{q} v-\beta R)^2}{2 x} - \Phi(y) \right].
\end{equation}
If we consider the pseudo--likelihood algorithm with $\Phi(y)=-\beta y + \ln 2 \cosh (\beta y)$
(see the the definition of $\Phi$ (\ref{eq:def_Phi}) and the cost function (\ref{pseudo-_cost})) we obtain the following equations for the order parameters:
\begin{subequations}
\begin{eqnarray}
&0 = \frac{1}{x}+\frac{\alpha \beta}{\sqrt{q}} \int \mathcal{D} v \; v \;  \left( 1- \tanh(\beta \hat{y}) \right) \label{eq:saddle_pl_1}\\
&0 = -\frac{R}{V x}+ \alpha \beta^2 \int \mathcal{D} v \left( 1- \tanh(\beta \hat{y}) \right) \label{eq:saddle_pl_2}\\
 &0 = -\frac{1}{x^2} \left( q-\frac{R^2}{V} \right) + \alpha \beta^2 \int \mathcal{D} v \; \left( 1- \tanh(\beta \hat{y}) \right) ^2 \label{eq:saddle_pl_3},
\end{eqnarray}
\end{subequations}
where $\hat{y}$ is defined by 
\begin{equation}
\hat{y}= \sqrt{q} v+\beta R + \beta x (1- \tanh(\beta \hat{y})).
\label{eq:y_sp}
\end{equation}

\section{Relation between order parameters}
\label{sec:AppB}
We introduce the auxiliary variables $ \{ \eta_1, \eta_2 \}$ 
in the free energy $F=F_0+F_1$ as follows:
\begin{equation}
\begin{split}
F_0(R,q,q_0,\eta_1, \eta_2) &= -   \lim_{n \to 0}  \nu^{-1}  N^{-1}  \frac{\partial}{\partial n} \ln  \int \prod_a d\bm{W^a}\; d \hat{q_0} \; d\hat{R} \; d\hat{q}  \\
&\prod_{a } e^{i\hat{R}(\sum_{ijk} U_{ij}  W^a_j  (\Lambda_i+\eta_1) U_{ik} W^*_k   - N R)} 
 \prod_{a } e^{i\hat{q_0}(\sum_{ijk} U_{ij}  W^a_j  \Lambda_i U_{ik} W^a_k  - N q_0)} \\
& \prod_{a < b} e^{i\hat{q}(\sum_{ijk} U_{ij}  W^a_j  (\Lambda_i+\eta_2) U_{ik} W^b_k  - Nq) }.
 \label{eq:Z_param}
\end{split}
\end{equation}
The integration gives
\begin{equation}
\begin{split}
F_0(R,q,q_0,\eta_1, \eta_2) =&\underset{\hat{R} ,\hat{q},  \hat{q_0}}{\mbox{extr}}
\;  \frac{1}{\nu} \left\{
i \hat{q_0} q_0 +i \hat{R} R - \frac{i}{2}   \hat{q} q \right. \\
&\left. + \frac{1}{2N} \sum_i 
\frac{ \hat{q} ( \Lambda_i+ \eta_2)  + i \hat{R} ^2  (\sum_j U_{ij} W^*_j)^2 ( \Lambda_i+ \eta_1) ^2}{ 2 \hat{q_0} - \hat{q}( \Lambda_i+ \eta_2) } \right. \\
&\left. +  \frac{1}{2N} \sum_i  \ln \left[ i\hat{q}  ( \Lambda_i+ \eta_2) - 2 i \hat{q_0}  \right] -  \frac{1}{2} \ln(2 \pi)\right\}.
\end{split}
\label{eq:G_relation_param}
\end{equation}
From  (\ref{eq:Z_param}) it is easy to see that 
the parameters  $\{ \rho,Q \}$ can be derived by  derivatives of the free energy:  
\begin{equation}
\begin{split}
&\rho= N^{-1} \overline{\bm{W^*} \cdot \langle  \bm{W} \rangle}=\frac{ \nu}{i \hat{R}} \frac{\partial F_0}{\partial \eta_1}, \\
&Q= N^{-1} \overline{\langle \bm{W}\rangle^2} =-\frac{2 \nu}{i \hat{q} }\frac{\partial  F_0}{\partial \eta_2}
\end{split}
\label{eq:relation_param}
\end{equation}
in the limit $\{ \eta_1 \to 0, \eta_2 \to 0 \}$, where $\hat{R}$ and  $\hat{q}$ are the values extremizing (\ref{eq:G_relation_param}) in the limit $\{ \eta_1 \to 0, \eta_2 \to 0 \}$: 
\begin{equation}
\begin{split}
& \hat{q}=i \frac{R^2 -Vq}{V (q_0-q)^2}, \\
& \hat{R}=i\frac{R}{V(q-q_0)},\\
& \hat{q_0}=i \frac{R^2+V(q_0-2 q)}{2 V (q_0-q)^2}.
\end{split}
\label{eq:saddle_eta}
\end{equation}
From  (\ref{eq:G_relation_param}), (\ref{eq:relation_param}) and (\ref{eq:saddle_eta}) we recover (\ref{eq:def_rho_Q}).

\section{Replica result for quadratic cost functions}
\label{sec:AppC}
If the cost function has the simple quadratic form (\ref{eq:quad_cost}), computing the maximum and the integrals in (\ref{eq:free_energy}) can be done analytically and the free energy is 
\begin{equation}
F= - \frac{q-R^2/V}{2x} + \alpha \frac{ q + \beta R (\beta R - 2 \eta)-\eta^2 x}{2(1 +  x)},
\label{eq:final_F_MF--ML}
\end{equation}
where the order parameters obey to the following saddle point equations:
\begin{equation}
\begin{split}
&q = \frac{\eta^2  (1+ \alpha \beta^2 V) }{(\alpha-1)  (1+\beta^2 V) },\\ 
&R = \frac{\eta \beta V}{1+\beta^2 V},\\
 &x = \frac{1}{(\alpha-1)}.
 \end{split}
\label{eq:saddle_MF--ML}
\end{equation}
With this result, the reconstruction error (\ref{eq:error_def_formula})
for the linear estimator becomes (\ref{eq:error_quad}).
Moreover, we can compute the parameter 
$\Delta$ defined in (\ref{eq:Delta}) as follows.
If the 'training energy' per degree of freedom $N$ becomes self--averaging
we can use the relation (see also (\ref{eq:replica_trick}))
\begin{equation}
E(\bm{W}^{ML}) = - \lim_{\nu\to\infty} \nu^{-1} \ln Z 
\label{eq:def_E}
\end{equation}
to explicitly evaluate  the minimum training energy as 
\begin{equation}
E(\bm{W}^{ML})= \frac{N}{\alpha} F(\bm{W}^{ML}) = - \frac{N \eta^2}{2 \alpha}  \frac{(1+\alpha \beta^2 V)}{(1+\beta^2 V)},
\label{eq:E_min} 
 \end{equation}
where the second equality follows from (\ref{eq:final_F_MF--ML}) with the order parameters fixed to their saddle point values
 (\ref{eq:saddle_MF--ML}).
From (\ref{E_phi}) and (\ref{eq:E_min}), one finds (\ref{delta_explicit}).

 \section{Asymptotics from the replica approach}
 \label{sec:AppD}
In the large $\alpha$ limit, we know that the parameter $x$ gets small and the parameters $q$ and $R$ both converge to $V$. For the pseudo--likelihood estimator we find the following relation, starting from (\ref{eq:saddle_pl_2}) in the limit $R \to V$ and (\ref{eq:saddle_pl_3}):
\begin{equation}
 q-\frac{R^2}{V} = \frac{1}{ \alpha \beta^2 } \frac{ \int \mathcal{D} v \; \left( 1- \tanh(\beta \hat{y}) \right) ^2 }{\left[ \int \mathcal{D} v \left( 1- \tanh(\beta \hat{y}) \right) \right]^2}
 \label{eq:asymp1}
\end{equation}
where $\hat{y}$ is given by (\ref{eq:y_sp}), that in the limit of small $x$ becomes $\hat{y} \simeq \sqrt{q} v+\beta R $. Via a change of variable, we find the following result in the limit $R \to V, q \to V$:
\begin{equation}
 \begin{split}
 q-\frac{R^2}{V} & = \frac{1}{ \alpha \beta^2 } \frac{ \int d v \; G_{\beta V,V} (v) \left( 1- \tanh(\beta v) \right) ^2 }{\left[  \int d v \; G_{\beta V,V} (v) \left( 1- \tanh(\beta v) \right) \right]^2} \\ 
 &\simeq 
 \frac{1}{ \alpha \beta^2 } \frac{1}{\int d v \; G_{\beta V,V} (v) \left( 1- \tanh^2(\beta v) \right) },
  \end{split}
\end{equation}
where in the last equality we exploited the relation $\int d v \; G_{\beta V,V} (v)  \tanh(\beta v) =\int d v \; G_{\beta V,V} (v)  \tanh^2(\beta v) $. Hence, the error (\ref{eq:error_def_formula}) for large $\alpha$ scales as 
\begin{equation}
\varepsilon \simeq \left(q - \frac{R^2}{V} \right)\frac{1}{N} \mbox{Tr}  \overline{C^{-1}}\simeq
\frac{1}{ \alpha \beta^2 } \frac{1}{\int d v \; G_{\beta V,V} (v) \left( 1- \tanh^2(\beta v) \right) }
\; \frac{1}{N} \mbox{Tr}  \overline{C^{-1}}.
\end{equation}

\section{Asymptotic error for pseudo--likelihood estimator}
\label{sec:AppE}
We show that $U_{ij} = - B_{ij}$ for the pseudo--likelihood case assuming  
that the model is matched to the true data generating distribution. We start from the relations
\begin{equation}
U_{ij}= 
\sum_{\bm{ \sigma}_{ \setminus 0}}  P(\bm{ \sigma}_{ \setminus 0}) \sum_{\sigma_0} P(\sigma_0\vert  \bm{ \sigma}_{ \setminus 0})\partial_i \partial_j \ln P(\sigma_0 \vert  \bm{ \sigma}_{ \setminus 0})
\end{equation}
and
\begin{equation}
B_{ij}= \sum_{\bm{ \sigma}_{ \setminus 0}}  P(\bm{ \sigma}_{ \setminus 0}) \sum_{\sigma_0} P(\sigma_0\vert  \bm{ \sigma}_{ \setminus 0})\partial_i \ln P
(\sigma_0 \vert  \bm{ \sigma}_{ \setminus 0})
\partial_j \ln P(\sigma_0 \vert  \bm{ \sigma}_{ \setminus 0}).
\end{equation}
We next perform the inner expectation over $P(\sigma_0\vert  \bm{ \sigma}_{ \setminus 0})$. The result follows
from 
$$
\partial_i \partial_j \ln P = - \partial_i \ln P\;  \partial_j \ln P + \frac{1}{P} \partial_i \partial_j P
$$
and  the fact that, by normalisation of $P$, one gets $\sum_{\sigma} \partial_i \partial_j P(\sigma \vert  \bm{ \sigma}_{ \setminus 0}) = 0$.

\section{Error  dependence on the system size}
\label{sec:AppF}
\begin{figure}[h!]

\centering
\vspace*{1\baselineskip}
\includegraphics[width=0.48\linewidth]{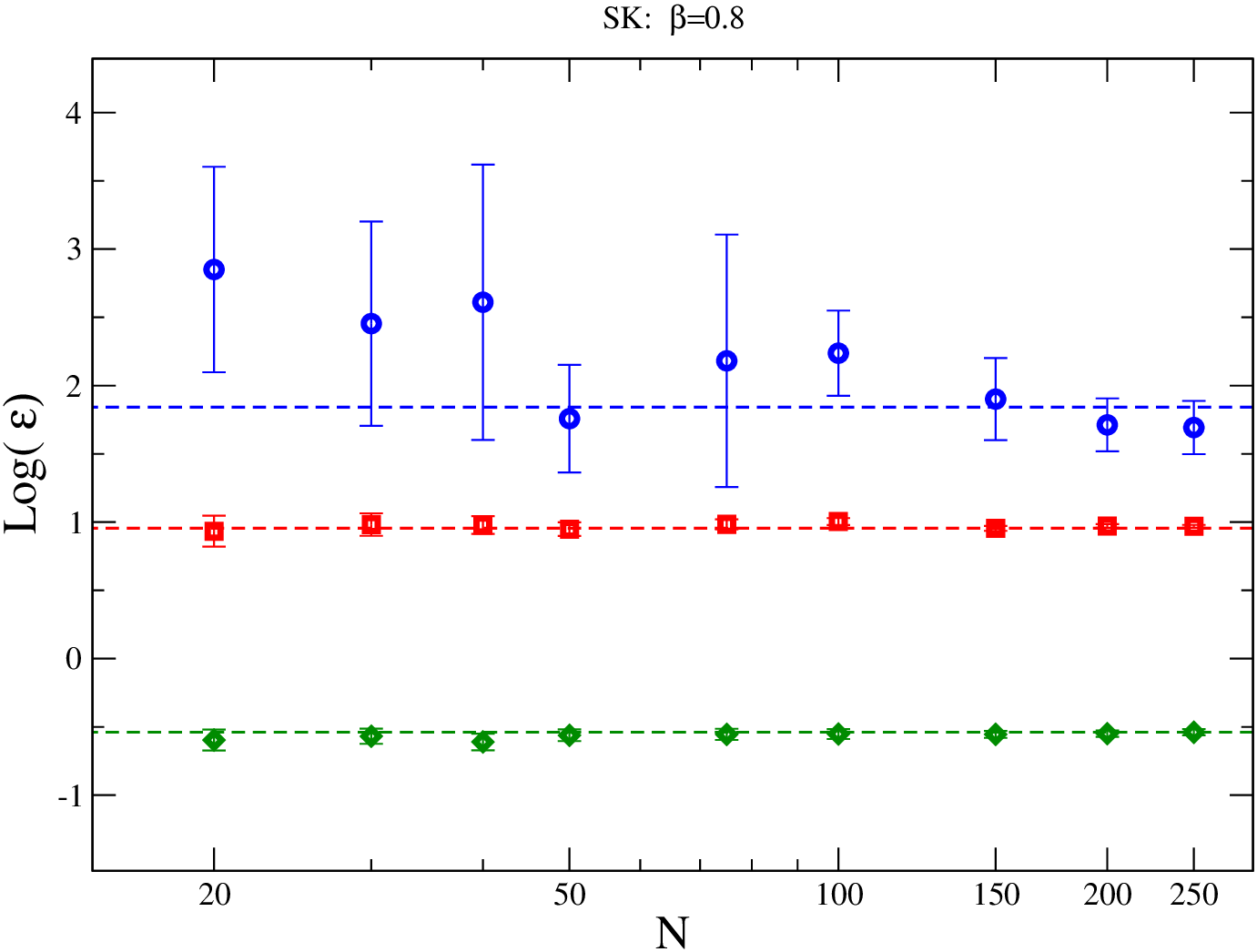} \hfill
\includegraphics[width=0.48\linewidth]{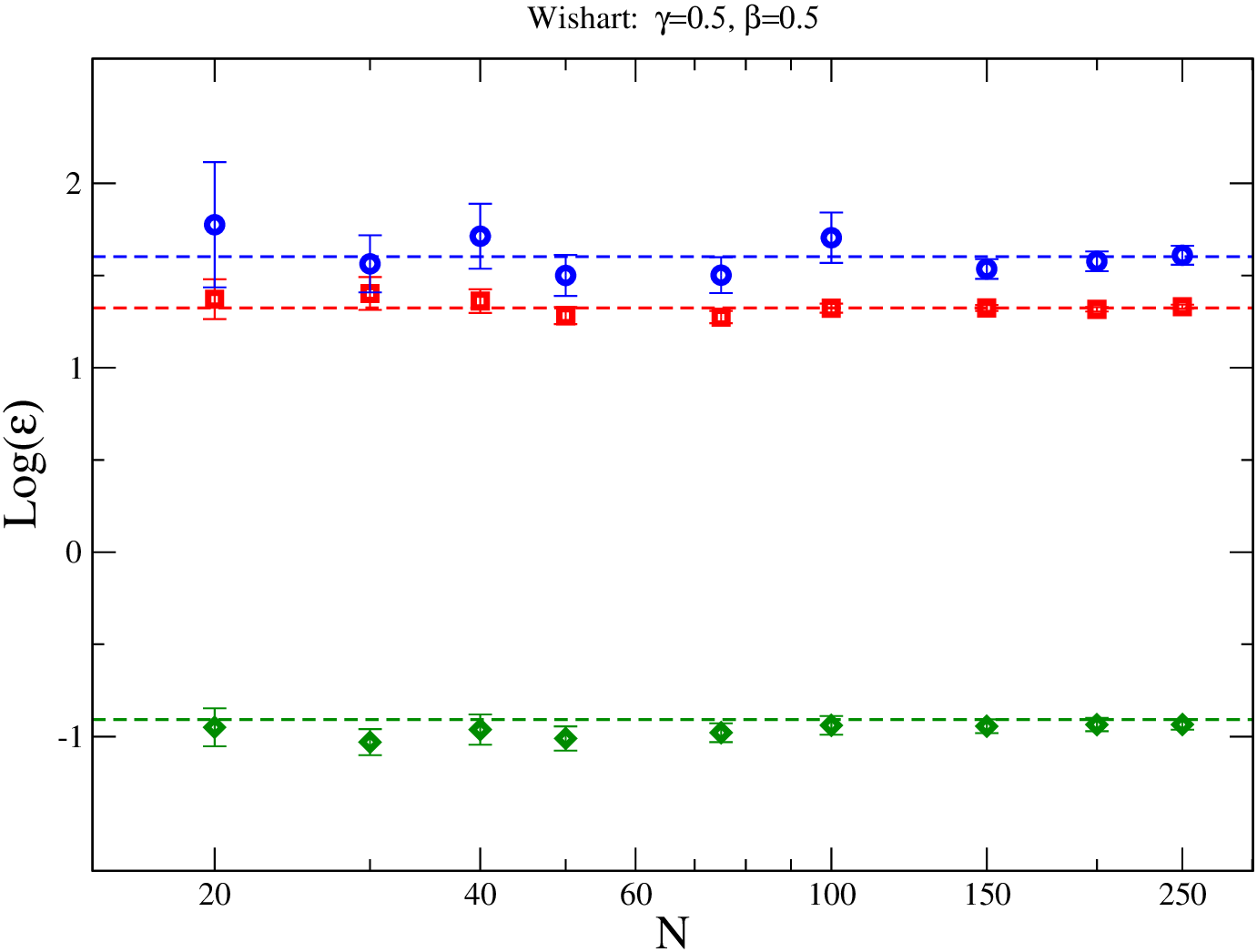}\\

\caption{Mean squared error of the couplings inferred by using the pseudo--likelihood algorithm (blue dots) the optimal local algorithm (green dots) and the MF--ML algorithm (red dots) as a function of the system size $N$, for fixed $\alpha=5$. The dotted lines represent replica results. Two different systems are considered: SK model at $\beta=0.8$ (left) and Wishart model with $\gamma=0.5$ at $\beta=0.5$ (right). The results are averaged over $5$ realizations of the network and $20$ different datasets generated from each network. Error bars represent standard deviations of the means. }
\label{fig:size}
\end{figure}

In Figure \ref{fig:error} we showed that the reconstruction error in systems with $N=100$ spins well agrees with the replica result, which is valid in the thermodynamic limit. However, it is relevant for applications to show an example of how  the system size  can affect the reconstruction error. In Figure \ref{fig:size} we show results obtained by fixing $\alpha$ and varying $N$. First of all we notice that finite size effects are much stronger for the the pseudo--likelihood algorithm than for the other two methods. Moreover, while the optimal local estimator always seems to outperform the other two methods, the performance difference between MF--ML and pseudo--likelihood algorithms  depends more strongly on the system parameters ($\alpha$, $\beta$, teacher coupling distribution), if $N$ is small. For instance, we see in Figure  \ref{fig:size} that, for systems of $N=20,30$ spins with couplings drawn from the Wishart distribution, the error of the MF--ML and pseudo--likelihood algorithms are compatible.

\end{appendices}

\section*{References}
\bibliographystyle{fbs}       
\bibliography{Bibliography}   
\end{document}